\documentclass[aps,nofootinbib,showpacs]{revtex4}

\usepackage{amsmath}
\usepackage{amssymb}
\usepackage{amsthm}
\usepackage{array}
\usepackage{float}
\usepackage{graphicx}
\usepackage{subfigure}

\begin{document}
\title{White Dwarf Pulsars as Possible Cosmic Ray Electron-Positron Factories}
\author{Kazumi Kashiyama}
\email{kashiyama@tap.scphys.kyoto-u.ac.jp}
\affiliation{Department of Physics,~Kyoto~University,~Kyoto~606-8502,~Japan}

\author{Kunihito Ioka}
\email{kunihito.ioka@kek.jp}
\author{Norita Kawanaka}
\email{norita@post.kek.jp}
\affiliation{Theory Center,~KEK(High~Energy~Accelerator~Research~Organization),~Tsukuba~305-0801,~Japan}

\begin{abstract}
We suggest that white dwarf~(WD) pulsars can compete with neutron star~(NS) pulsars
for producing the excesses of cosmic ray electrons and positrons 
($e^{\pm}$) observed by the PAMELA, ATIC/PPB-BETS, Fermi and H.E.S.S experiments.
A merger of two WDs leads to
a rapidly spinning WD 
with a rotational energy ($\sim 10^{50} \mathrm{erg}$) 
comparable to the NS case.
The birth rate ($\sim 10^{-2} \mbox{-} 10^{-3} \mathrm{/yr/galaxy}$) is also similar, 
providing the right energy budget for the cosmic ray $e^{\pm}$.
Applying the NS theory, we suggest that the WD pulsars
can in principle produce $e^{\pm}$ up to $\sim 10$ TeV.
In contrast to the NS model, the adiabatic and radiative energy losses of
$e^{\pm}$ are negligible since their injection continues 
after the expansion of the pulsar wind nebula,
and hence it is enough that
a fraction $\sim 1 \%$ of WDs are magnetized 
($\sim 10^7$--$10^9$ G) as observed.
The long activity also increases the number of nearby sources 
($\sim 100$), which reduces the Poisson fluctuation in the flux.
The WD pulsars could dominate the quickly cooling $e^{\pm}$ above TeV energy
as a second spectral bump
or even surpass the NS pulsars in the observing energy range $\sim 10\mathrm{GeV} \mbox{-} 1 \mathrm{TeV}$,
providing a background for the dark matter signals
and a nice target for the future AMS-02, CALET and CTA experiment.
\end{abstract}
\maketitle

\vspace{2mm}

\section{Introduction}\label{sec1}
Recently, the observational windows to 
the electron and positron ($e^{\pm}$) cosmic rays
are rapidly expanding the energy frontier,
revealing new aspects of our Universe.
The PAMELA satellite \cite{Adriani:2008zr}
shows that the cosmic ray positron fraction 
(the ratio of positrons to electrons plus positrons)
rises in the energy range of $10$ to $100$ GeV,
contrary to the theoretical prediction of secondary positrons
produced by hadronic cosmic rays interacting with the interstellar medium (ISM)~\cite{Moskalenko:1997gh}.
Shortly thereafter, ATIC/PPB-BETS \cite{chang:2008,Torii:2008xu} suggest an sharp excess of the $e^{\pm}$ with a peak at $600$ GeV,
and although not confirming the ATIC/PPB-BETS sharp peak spectrum~\footnote{The difference between the ATIC/PPB-BETS and Fermi results is still under debate \cite{Israel:2009}. In this paper we call these features as a whole "{\it excesses}".}, Fermi~\cite{Ackermann:2010_8,Abdo:2009zk,Moiseev:2007js} and H.E.S.S \cite{Collaboration:2008aa,Aharonian:2009ah} also suggest an excess of the $e^{\pm}$ total flux around $100$ GeV -- $1$ TeV compared to theoretical predictions 
based on low energy cosmic ray $e^{\pm}$ spectrum~\cite{Baltz:1998xv,Ptuskin:2006b}. 
All these observations of the $e^{\pm}$ excesses probably connected with the PAMELA positron excess, and most likely suggest a new source, possibly 
the astrophysical accelerators 
\cite{Kawanaka:2009dk,Hooper:2008kg,Yuksel:2008rf,Profumo:2008ms,Malyshev:2009tw,Grasso:2009ma,Kistler:2009wm,Heyl:2010md,Fujita:2009wk,Shaviv:2009bu,Hu:2009bc,Blasi:2009hv,Blasi:2009bd,Mertsch:2009ph,Biermann:2009qi,Ahlers:2009ae,Kachelriess:2010gt,Heinz:2002qj,Ioka:2008cv,Calvez:2010fd,KIOK:2010}
or dark matter annihilation 
\cite{Asano:2006nr,ArkaniHamed:2008qn,Baltz:1998xv,Barger:2008su,Barger:2009yt,Bergstrom:2008gr,Bertone:2008xr,Borriello:2009fa,Chen:2008fx,Cheng:2002ej,Cholis:2008qq,Cholis:2008wq,Cirelli:2008pk,Cirelli:2008jk,Crocker:2010gy,Feldman:2009wv,Fox:2008kb,Hall:2008qu,Harnik:2008uu,Hisano:2004ds,Hisano:2008ah,Hisano:2009rc,Hooper:2008kv,Hooper:2009fj,Feldman:2009,Ibe:2008ye,Ishiwata:2008cv,Kadota:2010xm,MarchRussell:2008tu,Meade:2009iu,Nomura:2008ru,Pospelov:2008jd,Yin:2008bs,Zavala:2009zr,Zhang:2008tb} 
/decay 
\cite{Arvanitaki:2009yb,Arvanitaki:2008hq,Barger:2009yt,Borriello:2009fa,Buchmuller:2009xv,Chen:2008dh,Chen:2008yi,Chen:2008qs,Cirelli:2008pk,DeLopeAmigo:2009dc,Fukuoka:2009cu,Hamaguchi:2008ta,Hamaguchi:2008rv,Hisano:2008ah,Ibarra:2008jk,Ibarra:2009dr,Ishiwata:2008cu,Mardon:2009gw,Meade:2009iu,Nardi:2008ix,Okada:2009bz,Shirai:2009fq,Yin:2008bs,Zhang:2008tb},
although there might remain alternatives such as
the propagation effects
\cite{Delahaye:2007fr,Cowsik:2009ga,Katz:2009yd,Stawarz:2009ig,Schlickeiser:2009qq}
or proton contamination \cite{Israel:2009,Fazely:2009jb,Schubnell:2009gk}.
These discoveries have excited 
the entire particle and astrophysics communities
and prompted over 300 papers within a year.
See~\cite{YiZhong:2010} for a recent review.

The most fascinating possibility for the $e^{\pm}$ excesses is the dark matter,
such as weakly interacting massive particles (WIMPs) 
that only appear beyond the Standard Model.
Dark matter is a stable particle that accounts most of the matter 
in the Universe but the nature is not known yet.
Usually, the observed $e^{\pm}$ excesses are far larger than expected in 
the conventional dark matter annihilation scenarios.
The annihilation cross section must be enhanced by 
two or three orders of magnitudes larger than that 
for dark matter to leave the desired thermal relic density.
Astrophysical boosts from substructure are
difficult to accommodate such large enhancements.
A possible solution is that dark matter interacts 
with a light force carrier, enhancing the annihilation 
by the Sommerfeld effect, only at the present time (not at freeze out)
\cite{ArkaniHamed:2008qn,Cholis:2008qq,Hisano:2004ds}.
The other possibilities include
the dark matter decay 
\cite{Arvanitaki:2009yb,Arvanitaki:2008hq,Barger:2009yt,Borriello:2009fa,Buchmuller:2009xv,Chen:2008dh,Chen:2008yi,Chen:2008qs,Cirelli:2008pk,DeLopeAmigo:2009dc,Fukuoka:2009cu,Hamaguchi:2008ta,Hamaguchi:2008rv,Hisano:2008ah,Ibarra:2008jk,Ibarra:2009dr,Ishiwata:2008cu,Mardon:2009gw,Meade:2009iu,Nardi:2008ix,Okada:2009bz,Shirai:2009fq,Yin:2008bs,Zhang:2008tb}
and the annihilation boosted by resonances \cite{Feldman:2009,Ibe:2008ye}.
Because the PAMELA anti-proton observations show no excess 
\cite{Adriani:2008zq,Adriani:2010rc},
any dark matter model should preferentially produce leptons rather than hadrons.
The other multi-messenger constraints
with radio, gamma-ray and neutrino observations
are also getting tight but not completely excluding 
the dark matter models
\cite{Meade:2009iu,Cirelli:2008pk,Yin:2008bs,Barger:2008su,Barger:2009yt,Nardi:2008ix,Zhang:2008tb,Bertone:2008xr,Zavala:2009zr,Borriello:2009fa,Crocker:2010gy,Hisano:2008ah,DeLopeAmigo:2009dc,Ishiwata:2008cu,Ackermann:2010rg,Abdo:2010ex,Abdo:2010dk}.

More conservative candidates are the astrophysical accelerators in our Galaxy, 
such as neutron star~(NS) pulsars \cite{Kawanaka:2009dk,Hooper:2008kg,Yuksel:2008rf,Profumo:2008ms,Malyshev:2009tw,Grasso:2009ma,Kistler:2009wm,Heyl:2010md}, 
supernova remnants~(SNRs)\cite{Fujita:2009wk,Shaviv:2009bu,Hu:2009bc,Blasi:2009hv,Blasi:2009bd,Mertsch:2009ph,Biermann:2009qi,Ahlers:2009ae,Kachelriess:2010gt}, 
microquasars \cite{Heinz:2002qj},
or possibly a gamma-ray burst \cite{Ioka:2008cv,Calvez:2010fd}.
Under plausible assumptions, they can supply sufficient energy 
for $e^{\pm}$ cosmic rays, as already known before the PAMELA era 
\cite{Shen70,MaoShen72,boulares89,Aha:95,Atoyan:1995,chi:1996,zhang:2001,grimani:2007,Buesching:2008hr,ShenBerkey68,cowsik79,erlykin02,pohl98,Moskalenko:1997gh,Strong:1998fr,Kobayashi:2003kp,Berezhko:2003pf,Strong:2004de}.
Cosmic ray $e^{\pm}$ propagate via diffusion in our Galaxy
deflected by magnetic fields \cite{Berezinski:1990}.
Since $e^{\pm}$ cannot propagate far away due to energy losses
by the synchrotron and inverse Compton emission,
the sources should be located nearby ($\lesssim 1$ kpc).
This proximity of the source provides a chance to directly probe the 
as-yet-unknown cosmic particle acceleration \cite{Kobayashi:2003kp} and 
investigate how the $e^{\pm}$ cosmic rays escape from the source to the ISM~\cite{KIOK:2010}.
Unlike dark matter, the astrophysical models generally predict, 
if at all, a broad spectral peak due to the finite source duration
\cite{Kawanaka:2009dk,Ioka:2008cv}.
The hadronic models such as SNRs
also predicts the antiproton excess above
$\sim 100$ GeV \cite{Fujita:2009wk,Blasi:2009bd}
(but see \cite{Kachelriess:2010gt}), 
as first pointed out by Fujita et al. \cite{Fujita:2009wk},
as well as the excesses of secondary nuclei such as 
the boron-to-carbon and
titanium-to-iron ratio \cite{Mertsch:2009ph,Ahlers:2009ae}.
The arrival anisotropy
\cite{MaoShen72,Buesching:2008hr,Ioka:2008cv}
is also useful to
discriminate between dark matter and astrophysical origins.
The exciting thing is that these signatures will be soon proved
by the next generation experiments,
such as AMS-02 \cite{Beischer:2009,Pato:2010ih}, 
CALET \cite{torii:2006,torii:2008}
onboard the Experiment Module of 
the International Space Station,
and CTA~\cite{CTA:2010} on the ground, in coming several years.

With the forthcoming next breakthrough,
it is important to lay down the theoretical foundation for
the TeV $e^{\pm}$ windows.
In particular, there could still be room for additional astrophysical signals
since the $e^{\pm}$ cosmic rays have only 
$\lesssim 1\%$ energy budget of the hadronic cosmic rays.
Although the supernova~(SN)-related sources such as NS pulsars and SNRs may be the most plausible sources of the TeV $e^{\pm}$, 
there should be only a few local sources~\cite{Watters:2010}, while $e^{\pm}$ from distant sources can not 
reach us due to the fast inverse Compton and synchrotron cooling  
\cite{Kobayashi:2003kp,Kawanaka:2009dk}.
Hence a clean window is possibly open for the dark matter or
other astrophysical signals.
A part of this window may have been already implied by
the spectral cutoff around $\sim 1$ TeV 
in the H.E.S.S. data.
The future AMS-02 experiment will detect $e^{\pm}$ 
up to $\sim 1$ TeV \cite{Beischer:2009,Pato:2010ih}, 
while CALET will observe electrons up to $\sim 10$ TeV 
with an energy resolution better than a few \% ($>100$ GeV) 
\cite{torii:2006,torii:2008}. Also CTA will be able to measure 
the cosmic ray electron spectrum up to $\sim 15$ TeV~\cite{CTA:2010}. 

In this paper, we propose yet another $e^{\pm}$ source
-- white dwarf~(WD) pulsars -- that
could potentially dominate the $\gtrsim$ TeV $e^{\pm}$ window or 
even already have been detected as the $e^{\pm}$ excesses
above the conventional models~\cite{Baltz:1998xv,Ptuskin:2006b}.
A WD pulsar is an analogue of the NS pulsar
with the central compact object being a WD
rather than a NS.
A spinning magnetized compact object generates huge
electric fields 
(potential differences) in the magnetosphere via unipolar induction
\cite{goldreich:1969,ruderman:1975,cheng:1986},
and accelerates particles to produce $e^{\pm}$ pairs
if certain conditions are met.
Then, almost all the spindown energy
is transferred to the outflows of relativistic $e^{\pm}$,
resulting in the cosmic ray $e^{\pm}$.

In our model, a rapidly spinning WD
is mainly formed by a merger of two ordinary WDs
(or possibly by an accretion),
since the observed WDs are usually slow rotators
\cite{Kawaler:2003sr}.
Such a merger scenario was proposed to explain Type Ia supernovae~(SNIa).
However, it is not clear that such mergers lead to 
the SN explosions \cite{Pakmor:2009yx}.
It seems reasonable that
about half of mergers leave rapidly spinning WDs
with the event rate of about one per century in our Galaxy
\cite{Nelemans:2001hp,Farmer:2003pa}.
The strong magnetic fields ($> 10^6$ G) are also expected
as a fraction $\sim 10\%$ of WDs~\cite{Liebert:2002qu,Schmidt:2003ip}.
Combining these facts, we will estimate 
that the WD pulsars
can potentially provide the right amount of energy 
for the cosmic ray $e^{\pm}$ (see Sec.\ref{sec2}).
We note that 
the WD mergers are also related to
the low frequency gravitational wave background for LISA~\cite{LISA:2010}.

The WD pulsars
have been theoretically adopted to interpret the observational features of the anomalous X-ray pulsars
\cite{paczynski:1990,usov:1993,usov:1988},
the close binary AE Aquarii \cite{Ikhsanov:2005qf},
and the transient radio source GCRT J1745--3009 \cite{Zhang:2005kz}.
Our calculations for the $e^{\pm}$ production
are essentially similar to that of Usov \cite{usov:1988,usov:1993}
and Zhang \& Gil \cite{Zhang:2005kz}.
However, this is the first time to apply the WD pulsars
to the $e^{\pm}$ cosmic rays, as far as we know.
We also discuss the adiabatic energy losses of
$e^{\pm}$ in the pulsar wind nebula, that are found to be negligible
in contrast to the NS model.
From the observational viewpoint,
the WD pulsars have not been firmly established,
whereas there are several indications for their existence,
such as the hard X-ray pulsation in AE Aquarii \cite{Terada:2007br}.
The WD pulsars are likely still
below the current level of detection
because they are rare, $\sim 10^{-4}$ of all WDs,
and relatively dim.

This paper is organized as follows.
In Sec.\ref{sec2}, we show that the WD pulsars can produce and accelerate $e^{\pm}$ up to the 
energy above TeV. At first we show that the energy budgets of WD pulsars are large enough to explain the PAMELA positron excess by order-of-magnitude estimates. Then we discuss, more closely, whether or not WD pulsars can produce and accelerate $e^{\pm}$ up to the energy above TeV by considering the magnetospheres and pulsar wind nebulae. We also point out that there should be much 
more nearby active WD pulsars compared with NS pulsars since the lifetime of WD pulsars are much longer. In Sec.\ref{sec3}, we discuss the propagation of the $e^{\pm}$ from WD pulsars, and 
show the possible energy spectrum observed by the current and future observations in the WD pulsar dominant model and 
the WD and NS pulsar mixed model.
As complements, we also give a short review of the current status of the observations of WD pulsar candidates. 
In Sec.\ref{sec4}, we summarize our paper and discuss open issues. 

\section{White dwarf pulsars}\label{sec2}

\subsection{Energy Budgets of White Dwarf Pulsars}

In this subsection, we show that WDs potentially have enough rotational energy for producing high energy $e^{\pm}$ cosmic rays.

NS pulsars, which are formed after the SN explosions, are one of the most promising candidates for the astrophysical sources of 
high energy positrons. For the PAMELA positron excess, each NS pulsar should provide mean energy $\sim 10^{48}\ {\rm erg}$ to positrons~\cite{Kawanaka:2009dk,Hooper:2008kg}, since the energy budgets of cosmic ray positrons is $\sim 0.1\%$ of that of cosmic ray protons, which is estimated as $\sim 10^{50}\ {\rm erg}$ per each SN, and the positrons suffer from the radiative cooling during the propagation more than the protons.
The intrinsic energy source is the rotational energy of a newborn NS, which is typically
\begin{equation}\label{Erot1}
E_{\text{rot,NS}} \approx \frac{1}{2} I \Omega^2 \sim 10^{50} \left(\frac{M}{1.0 M_{\odot}} \right) \left(\frac{R}{10^{6} \text{cm}} \right)^{2} \left(\frac{\Omega}{10^{2} \text{s}^{-1}} \right)^{2} \text{erg},
\end{equation}
where $I$ is the moment of inertia of the NS. Then, if all the NS pulsars are born with the above rotational energy and the $\sim 1 \%$ energy is used 
for producing and accelerating $e^{\pm}$, the NS pulsars can supply enough amounts of $e^{\pm}$ for explaining the PAMELA positron 
excess~\cite{Kawanaka:2009dk}. 

Let us show that double degenerate WD binary mergers can also supply enough amounts of rotational energy. Here we consider the mass $0.6 
M_{\odot}$ and radius $R \sim 10^{8.7} \text{cm}$ for each WDs, 
which are typically observed ones~\cite{Falcon:2010}. Just after a merger of 
the binary, the rotational speed $v_{\text{rot}}$ can be estimated as $v_{\text{rot}} \approx (GM/R)^{1/2} \sim 10^8 \text{cm/s}$, which corresponds to the mass 
shedding limit, and the angular frequency is about $\Omega = v_{\text{rot}}/R \sim 0.1 \text{s}^{-1}$. Then, the rotation energy of the merged object is 
\begin{equation}\label{Erot2}
E_{\text{rot,WD}}\approx \frac{1}{2} I \Omega^2 \sim 10^{50}\left(\frac{M}{1.0 M_{\odot}} \right) \left(\frac{R}{10^{8.7} \text{cm}} \right)^{2} \left(\frac{\Omega}{0.1 \text{s}^{-1}} \right)^{2} \text{erg},
\end{equation}
which is comparable to the NS pulsar case in Eq.(\ref{Erot1}).
The event rate $\eta_{\text{WD}}$ of the double degenerate WD mergers in our Galaxy remains uncertain. Any theoretical estimate requires a knowledge 
of the initial mass function for binary stars, the distribution of their initial separation,
and also the evolution of the system during periods of nonconservative mass transfer. There are still reasonable estimates in the range~\cite{Nelemans:2001hp,Farmer:2003pa},
\begin{equation}\label{eta}
\eta_{\text{WD}} \sim 10^{-2}\mbox{--}10^{-3} \ /\text{yr}/\text{galaxy}.
\end{equation}
This is comparable to the typical birth rate of NS pulsars~\cite{Narayan:1987,Lorimer:1993}. Therefore, from the viewpoint of energy 
budget in Eqs. (\ref{Erot1}), (\ref{Erot2}) and (\ref{eta}), the WDs are also good candidates for the high energy $e^{\pm}$ sources as the 
NS pulsars, if the merged binaries can efficiently produce and accelerate $e^{\pm}$. 

The estimated merger rate is also similar to that of SNIa, which is one of the reason that the double degenerate WD 
mergers are possible candidates for SNIa. Since the typical WD mass is $0.6M_{\odot}$, the merged objects do not exceed the Chandrasekhar 
limit $1.4M_{\odot}$ even without any mass loss. Then, they leave fast rotating WDs, as suggested by some recent simulations~\cite{Loren:2009}, and could become WD pulsars. In this paper, we assume that a fair fraction of double degenerate WD mergers result in the WD pulsars.
\footnote{Since the highly magnetized WDs have higher mean mass
$\sim 0.95 M_{\odot}$ than the total average $\sim 0.6 M_{\odot}$~\cite{Liebert:2002qu},
the fraction of mergers that leave spinning WDs could be lower than the average.}

The accretion scenario is another possibility for the fastly rotating WD formation. In the single degenerate binary, which consists of a 
WD and a main sequence star, there should be a mass transfer from the main sequence star to the WD as the binary separation becomes 
smaller and the Roche radius becomes larger than the radius of the main sequence star. In this stage, the angular momentum is also transferred to the 
WD, and the WD can spin up to around the mass shedding limit with the rotational energy as large as Eq.(\ref{Erot2}). 
In Sec.\ref{observation}, we refer to such a WD pulsar candidate, AE Aquarii. 

Since the birth rate is relatively uncertain in the accretion scenario, we just concentrate on the merger scenario in this paper.

\subsection{$e^{\pm}$ Production and Acceleration}\label{Sec.condition}
In this subsection we discuss the possibility that WD pulsars emit high energy $e^{\pm}$ above TeV. 
In order to produce the TeV $e^{\pm}$, a pulsar has to 
\begin{enumerate}
\item[(i)]{produce $e^{\pm}$ pairs} 
\item[(ii)]{accelerate $e^{\pm}$ up to TeV.} 
\end{enumerate}
We show that WD pulsars can meet both of the conditions. From now on we set fiducial parameters of the WD pulsar's surface dipole magnetic field, 
angular frequency, and radius as $B_{\text{p}} = 10^{8}\text{G}$, $\Omega = 0.1 \text{s}^{-1}$ and $R = 10^{8.7}\text{cm}$, respectively.
For comparison, we set fiducial parameters of the NS pulsars as $B_{\text{p}} = 10^{12}\text{G}$, $\Omega = 10^2\text{s}^{-1}$ and $R = 10^{6}\text{cm}$.

\subsubsection{$e^{\pm}$ pair production in magnetosphere}

Some of the observed WDs have strong magnetic fields of $B \sim 10^{7\mbox{-}9} \text{G}$~\cite{Liebert:2002qu,Schmidt:2003ip}. For such 
WDs, if they are rapidly rotating as we discuss in the previous subsection, the electric field along the magnetic field are induced on the 
surface and the charged particles are coming out from the surface layer of the pulsars. Then we can expect that, as in the case of ordinary NS pulsars, the corotating magnetosphere are formed around the WDs, in which the charge distribution of plasma should be the 
Goldreichi-Julian (GJ) density in a stationary case~\cite{goldreich:1969},
\begin{equation}\label{GJ}
\rho_0 = {\bf \nabla} \cdot \frac{(\bf{\Omega} \times {\bf r}) \times {\bf B}}{4 \pi c} \approx - \frac{\bf{\Omega} \cdot \bf B}{2 \pi c} \sim -\frac{10^{5}}{|Z|} \left( \frac{B_{\text{p}}}{10^8 \text{G}}\right) \left( \frac{\Omega}{0.1\text{s}^{-1}}\right) \text{cm}^{-3},
\end{equation}
where $Z$ is the elementary charge of particles in the plasma. Here we assume that the large scale configuration of the magnetic field is dipole. Since the corotating speed of the magnetic field lines cannot exceed the speed of light, the magnetic field cannot be closed 
outside the light cylinder $R_{\text{lc}} = c/\Omega$. This fact leads to the open magnetic field lines in the polar region. The electric potential 
difference across this open field lines is~\cite{goldreich:1969}
\begin{equation}\label{DelVmax}
\Delta V_{\text{max}} = \frac{B_{\text{p}} \Omega^2 R^3}{2 c^2} \sim 10^{13} \left( \frac{B_{\text{p}}}{10^8 \text{G}}\right) \left( \frac{\Omega}{0.1 \text{s}^{-1}}\right)^{2} \left( \frac{R}{10^{8.7} \text{cm}}\right)^{3} \text{Volt},
\end{equation}
which is the maximum value for the pulsars in principle. 

If the GJ density is completely realized in the magnetosphere, electric fields along the magnetic field lines is absent: ${\bf E} \cdot {\bf B} = 0$
. Since the charged particles are tied to the strong magnetic field, the acceleration of $e^{\pm}$ cannot occur. That leads to the absence of 
high energy $\gamma$ ray emissions from the accelerated $e^{\pm}$ and successive pair production avalanches. However, there are two prospective 
scenarios of forming the region where the charge density is not equal to the GJ density, and hence $e^{\pm}$ are accelerated and produced in the 
NS pulsar magnetosphere, that is the polar cap~\cite{ruderman:1975,Arons:1979} and outer gap model~\cite{cheng:1986}. From now on, 
we assume that the magnetosphere structure of WD pulsars are similar to that of NS pulsars, and discuss the $e^{\pm}$ 
pair production especially in the polar cap region. 

In polar cap models, electric potential drops along the magnetic fields are formed in the polar region of the pulsars. There are some different 
types of polar cap models. First, the angle between the magnetic and rotational axis determines the sign of electric charge of the 
particles propagating along the open magnetic field lines in accordance to Eq.(\ref{GJ})~\cite{goldreich:1969}. The GJ density in the polar cap 
region is positive when ${\bf \Omega} \cdot {\bf B} < 0$ and negative when ${\bf \Omega} \cdot {\bf B} > 0$. Second, polar cap models depend on 
whether or not steady charge currents flow out from the surface of the pole region. After the GJ density is realized, there are no electric forces 
working on the charged particles in the surface layer. Hence, whether or not the charged particles come out from the surface is 
determined by the competition between the binding energy of ions or electrons at the surface and thermal energy. In the original model 
proposed by Ruderman and Sutherland~\cite{ruderman:1975}, they assume that the binding energy is bigger. Then due to the outflow along the 
open magnetic field, a gap where the charge density is almost $0$ is formed in the pole region. On the other hand, if the thermal energy is bigger, 
there exist a positive or negative space-charge-limited flow~\cite{Arons:1979}. Even in this case, it is shown that, by virtue of the 
curvature of magnetic fields, the charge density deviates from the GJ density and electric potential drops along the open magnetic field lines 
can be formed~\cite{Arons:1979}. Although a general relativistic frame dragging effect also contributes to form 
electric potential drops in the polar cap region~\cite{MuslimovTsygan:1991}, the effect can be neglected compared with the effect of magnetic field 
curvature in the case of the WDs~\cite{Zhang:2005kz}. 

In the polar cap region, where the GJ density is not realized, primary electrons or positrons are accelerated, and they emit curvature radiations, 
which interact with the magnetic fields and produce secondary $e^{\pm}$ pairs, $\gamma + B \rightarrow e^- + e^+ $~\cite{ruderman:1975}. The 
secondary $e^{\pm}$ are also accelerated and emit curvature radiations 
that produce further $e^{\pm}$ pairs~(pair creation avalanche). Inverse Compton scatterings can also serve as a way to produce high energy 
$e^{\pm}$ and successive pair creation avalanches~\cite{ZhangQiao:1996}. Due to the abundant charges supplied by the avalanche, the GJ density is 
realized at a finite distance from the surface and the polar cap formation stops. In the quasi-steady state, the size of the polar cap region can be 
approximated as $h \approx l$, where $l$ is the mean free path of the $e^{\pm}$ pair creation process. To put it the other way around, only when the 
available size of the polar cap region $h_{\text{max}}$ is larger than $l$, $e^{\pm }$ pair creation avalanches can be formed. Chen \& Ruderman first 
derived the condition for NS pulsars and succeeded in showing the NS pulsar "death line"~\cite{ChenRuderman:1993}. Harding \& Muslimov also derived 
the NS death line under more general conditions~\cite{Harding:2001,Harding:2002}. Here we follow Chen \& Ruderman's approach and drive the $e^{\pm}$ 
pair production avalanche condition in the case of WD pulsars. We discuss the validity of this simple treatment in Sec.\ref{sec4}.

Going through any potential drop $\Delta V$ along the open magnetic field lines, $e^{\pm}$ are accelerated up to the Lorentz factor  
\begin{equation}\label{gamma}
\gamma = \frac{e \Delta V }{m_{\text{e}} c^2},
\end{equation}
where $m_{\text{e}}$ is the mass of electrons. The characteristic frequency of curvature radiation photons from the accelerated $e^{\pm}$ is
\begin{equation}\label{omega}
\omega_{\text{c}} = \gamma^3 \frac{c}{r_{\text{c}}},
\end{equation}
where $r_{\text{c}}$ is the curvature radius of the magnetic field lines. The mean free path of a photon of energy $\hbar \omega > 2m_{\text{e}} c^2$ moving through a 
region of magnetic fields is~\cite{Erber:1966}
\begin{equation}
l = 4.4 \ \frac{\hbar c}{e^2}\frac{\hbar}{m_{\text{e}} c} \frac{B_{\text{q}}}{B_{\perp}} \exp \left( \frac{4}{3 \chi} \right); \ (\chi << 1), \notag
\end{equation}
\begin{equation}\label{chi}
\chi\equiv \frac{\hbar \omega}{2 m_{\text{e}} c^2} \frac{B_{\perp}}{B_{\text{q}}}.
\end{equation}
Here $B_{\text{q}}=m_{\text{e}}^2 c^3/e \hbar = 4.4 \times 10^{13}\text{G}$ and $B_{\perp} = B_{\text{s}} \sin \theta$ with $\theta$ is the angle between the direction of 
propagation for photon and the surface magnetic field lines of the pulsars. $B_{\text{s}}$ is the local magnetic field at the surface of the pulsar which is 
not necessarily coincident with the dipole field $B_{\text{p}}$. At distance $h$ above the pulsar surface, the $\sin \theta$ can be approximated to 
$\approx h/r_{\text{c}}$, then 
\begin{equation}\label{Bperp}
B_{\perp} \approx B_{\text{s}} \frac{h}{r_{\text{c}}}.
\end{equation}
We shall consider the situation $l \approx h$, however which could be realized when $\chi^{-1} = {\it O}(10)$ without relying on precise value of the 
parameters characterizing NS or WD pulsars since small changes in $\chi$ correspond to the exponentially large change in $l$. Here 
we take the critical value as $\chi^{-1} = 15$ following \cite{ChenRuderman:1993}. Substituting Eq.(\ref{gamma}), (\ref{omega}), (\ref{Bperp}) to 
Eq.(\ref{chi}), this condition is given by 
\begin{equation}\label{pair_cre_con}
\left( \frac{e \Delta V}{m_{\text{e}} c^2} \right)^3 \frac{\hbar}{2 m_{\text{e}} c r_{\text{c}}} \frac{h}{r_{\text{c}}} \frac{B_{\text{s}}}{B_{\text{q}}} \approx \frac{1}{15}.
\end{equation}
Eq.(\ref{pair_cre_con}) corresponds to a general condition for $e^{\pm}$ pair production avalanches in the polar cap region of pulsars. Then we 
have to specify $h$, $\Delta V$, $B_{\text{s}}$ and $r_{\text{c}}$. The thickness $h$ and the potential drop $\Delta V$ in the polar cap region depend on which polar 
cap model we adopt. Here we consider the original polar cap model proposed by Ruderman and Sutherland~\cite{ruderman:1975}. In this case, 
the relation between $h$ and $\Delta V$ is given by
\begin{equation}\label{DelV}
\Delta V = \frac{B_{\text{s}} \Omega h^2}{2c}.
\end{equation}
Then, since $\Delta V$ cannot exceed the maximum potential drop available in a pulsar magnetosphere, $\Delta V_{\text{max}}$ in Eq.($\ref{DelVmax}$), $h$ 
also cannot exceed the maximum thickness 
\begin{equation}
h_{\text{max}} \approx \left( \frac{R^3 \Omega}{c} \right)^{1/2}.
\end{equation}
$B_{\text{s}}$ and $r_{\text{c}}$ depend on the configuration of the surface magnetic field, which is very uncertain even in the case of the NS pulsars. 
Here we suppose curved magnetic fields in the polar cap region and set $r_{\text{c}} \approx R$ and  $B_{\text{s}} \approx B_{\text{p}}$. In this case, the 
condition for $e^{\pm}$ pair production avalanche (Eq.(\ref{pair_cre_con})) is   
\begin{equation}
\left( \frac{e \Delta V_{\text{max}}}{m_{\text{e}} c^2} \right)^3 \frac{\hbar}{2 m_{\text{e}} c R} \frac{h_{\text{max}}}{R} \frac{B_{\text{p}}}{B_{\text{q}}} \gtrsim \frac{1}{15},
\end{equation}
which is equivalent to
\begin{equation}\label{avalanche_polar}
4 \log B_{\text{p}} -6.5 \log P + 9.5 \log R \gtrsim 96.7,
\end{equation}
where the unit of $B_{\text{p}}$, $P=2\pi/\Omega$ and $R$ are [G], [sec] and [cm], respectively. By substituting $R \sim 10^{6}\text{cm}$, which is the typical 
radius of NSs, Chen and Ruderman succeeded in explaining the NS pulsar death line~\cite{ChenRuderman:1993}. In the case of 
WD pulsars, substituting our fiducial parameters $B_{\text{p}} \sim 10^8 \text{G}$, $P \sim 50 \text{s}$~($\Omega \sim 0.1 \text{s}$) and 
$R \sim 10^{8.7} \text{cm}$, we find that the WD pulsars well satisfy Eq.(\ref{avalanche_polar}), and thus also the condition (i) 
in Sec.\ref{Sec.condition}.

\begin{figure}[h]
\begin{center}
\includegraphics[width=110mm]{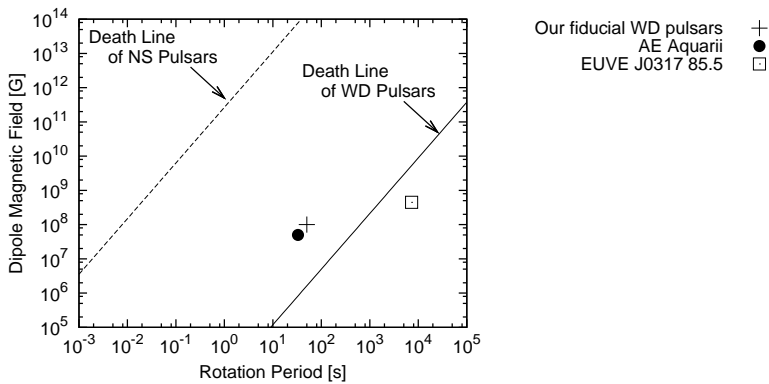}
\caption{This figure shows the death lines of WD (solid line) and NS (dashed line) pulsars. The cross shape indicates the fiducial 
parameters of the WD pulsars, $B_{\text{p}} = 10^8 \text{G}$ and $P = 50 \text{s}$. The observed data of rapidly rotating magentised WDs, AE 
Aquarii (filled circle) and EUVE J0317 855 (open square) are also 
plotted. The parameters and observational properties of these WDs 
are given in Sec.\ref{observation}.}\label{death_line}
\end{center}
\end{figure}

Fig.\ref{death_line} shows the death lines of the WD and NS pulsar with the fiducial parameters of the WD pulsars. We also plot 
parameters of the observed WD pulsar candidates, AE Aquarii and EUVE J0317. As we discuss in Sec.\ref{observation}, the pulse emission like ordinary 
NS pulsars are observed for AE Aquarii, and not for EUVE J0317, which is consistent with the death line. 

\subsubsection{$e^{\pm}$ acceleration and cooling in pulsar wind nebula}
In the previous subsection we show that WD pulsars can produce $e^{\pm}$ pairs in the magnetospheres. In this subsection we discuss the 
acceleration and cooling of the $e^{\pm}$ in the pulsar wind nebulae. 

\begin{figure}[h]
\begin{center}
\includegraphics[width=70mm, angle=-90]{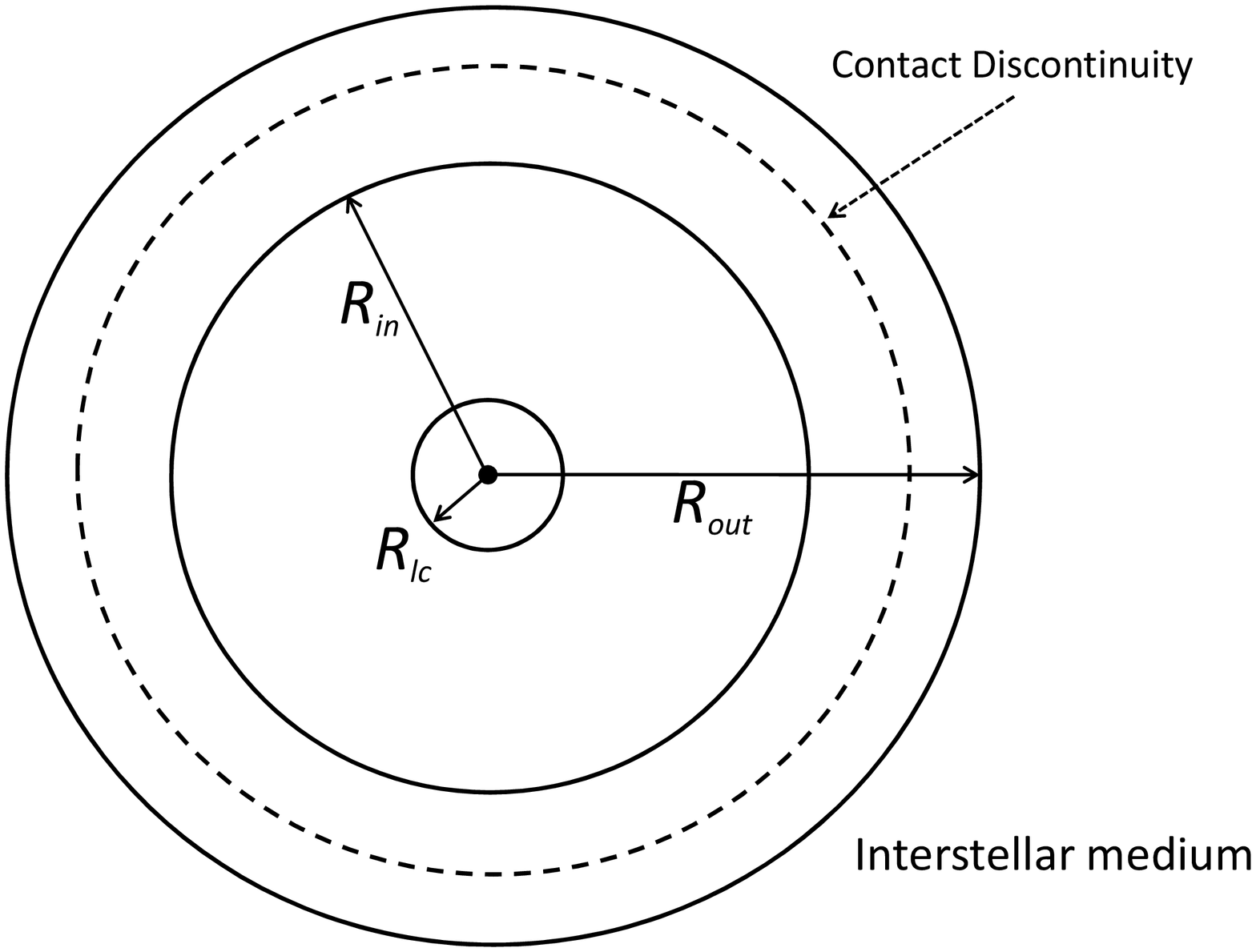}
\caption{This figure shows the schematic picture of the expected WD pulsar wind nebula.}\label{wind_nebula}
\end{center}
\end{figure}
Fig.\ref{wind_nebula} shows the schematic picture of a expected WD pulsar wind nebula. Once a WD pulsar is formed, the relativistic wind blasts 
off from the pulsar magnetosphere $\sim R_{\text{lc}}$. The supersonic wind becomes subsonic by passing the shock front at $\sim R_{\text{in}}$, reaches the ISM and 
forms a contact discontinuity. Since the wind is continuously injected by the pulsar, the contact discontinuity keeps sweeping the interstellar 
matter, and then the outer shock front is formed at $\sim R_{\text{out}}$. We emphasize that the SN shock front does not exist outside the shocked region 
unlike the NS pulsars since there suppose to be no SN explosion when the WD pulsar is formed. 

First we estimate the energy of $e^{\pm}$ available in the wind region $R_{\text{lc}} < r < R_{\text{in}}$. In principle, $e^{\pm}$ can be accelerated 
to the energy that the equipartition is realized between the wind and magnetic field, $\epsilon N = B^2/8\pi$
, that is
\begin{equation}\label{wind_gamma}
\epsilon = \frac{B^2}{8\pi N},
\end{equation}
where $N$ is the number density of $e^{\pm}$. If the number flux is conserved in the wind region, $4\pi r^2 c N \approx \text{const}$, $N$ can be 
described as
\begin{equation}
N = N_{\text{lc}} \left( \frac{R_{\text{lc}}}{r} \right)^2 ,
\end{equation}
where $N_{\text{lc}}$ is the number density at the light cylinder which can be estimated as  
\begin{equation}\label{Ndensity}
N_{\text{lc}} = \frac{\rho_{\text{lc}}}{e} {\cal M} = \frac{B_{\text{lc}}\Omega}{2\pi c e} {\cal M},
\end{equation}
where $\rho_{\text{lc}}$ and $B_{\text{lc}}$ are the GJ density (Eq.(\ref{GJ})) and magnetic field strength at the light cylinder, respectively and ${\cal M}$ is the 
multiplicity of $e^{\pm}$ in the magnetosphere. Inside the light cylinder $r < R_{\text{lc}} = c/\Omega$, the magnetic field is almost pure dipole, 
\begin{equation}\label{Bconf1}
B = B_{\text{p}} \left( \frac{R}{r} \right)^3 .
\end{equation}
For the fiducial parameters of WD pulsars, the radius of the light cylinder is $R_{\text{lc}} \sim 3\times 10^{11} \text{cm}$ and 
$B_{\text{lc}} = B_{\text{p}} (R/(c/ \Omega))^3 \sim 1 \text{G}$. Outside the light cylinder $r > R_{\text{lc}}$, if the energy flux of the magnetic field is also conserved 
$B\cdot r \approx \text{const}$, then
\begin{equation}\label{Bconf2}
B = B_{\text{lc}} \frac{R_{\text{lc}}}{r}. 
\end{equation}
Substituting $B_{\text{lc}} = B_{\text{p}} (\Omega R/c)^3$ and Eq.(\ref{Ndensity}) to Eq.(\ref{wind_gamma}), the typical energy of $e^{\pm}$ in the wind region can be 
described as 
\begin{equation}\label{emax}
\epsilon = \frac{e \Delta V_{\text{max}}}{{\cal M}} \sim 10 {\cal M}^{-1} \left( \frac{B_{\text{p}}}{10^{8}\text{G}} \right) \left( \frac{\Omega}{0.1 \text{s}^{-1}} \right)^2 \left( \frac{R}{10^{8.7}\text{cm}} \right)^3 \text{TeV},
\end{equation}
where $\Delta V_{\text{max}}$ is shown in Eq.(\ref{DelVmax}). The multiplicity of $e^{\pm}$ in the pulsar magnetosphere and wind nebula have not been understood 
clearly even in the case of NS pulsars and there are several discussions~\cite{chi:1996,zhang:2001}. Although details of 
the multiplicity in the magnetosphere cannot be discussed at this stage\footnote{In \cite{usov:1993}, Usov discussed the multiplicity in the magnetosphere for a X-ray pulsar 1E 2259+586 based on the WD pulsar model by investigating the observed X-ray luminosity, in which ${\cal M} \sim 0.1$.}, TeV energy $e^{\pm}$ could come out of the wind region and the condition (ii) in Sec.\ref{Sec.condition} can be 
fulfilled if ${\cal M}$ is not large.

Secondly we estimate the adiabatic and radiative cooling of $e^{\pm}$ in the shocked region. To that end, we have to identify the radii of the inner and outer shock front $R_{\text{in}}$ and $R_{\text{out}}$. The equation of motion for the outer shock front is
\begin{equation}\label{windEOM}
\frac{d}{dt} \left\{ \frac{4 \pi}{3} R_{\text{out}}^3 \rho \frac{dR_{\text{out}}}{dt} \right\} = 4\pi R_{\text{out}}{}^2 P_{\text{sh}},
\end{equation}
where $P_{\text{sh}}$ is the pressure of the shocked region and $\rho$ is the density of the ISM $\rho \sim 10^{-24} \text{g cm}^{-3}$. The 
energy conservation law at the outer shock front is
\begin{equation}\label{windEcon}
\frac{d}{dt} \left\{ \frac{4 \pi}{3} R_{\text{out}}{}^3 \frac{3}{2} P_{\text{sh}} \right\} = L - P_{\text{sh}} \frac{d}{dt} \left\{ \frac{4 \pi}{3} R_{\text{out}}{}^3 \right\} .
\end{equation}
Here $L$ is the spin down luminosity of WD pulsars,
\begin{equation}\label{Lspindown}
L = \frac{B_{\text{p}}^2 \Omega^4 R^6}{c^3},
\end{equation}
and we suppose that in the shocked region the particles are relativistic and its internal energy is $3P/2 $. 
Solving Eq.(\ref{windEOM}) and (\ref{windEcon}) for $R_{\text{out}}(t)$, 
\begin{equation}\label{Rout}
\begin{split}
R_{\text{out}}(t) &= \left( \frac{125}{154 \pi} \right)^{1/5} \left( \frac{L}{\rho} \right)^{1/5} t^{3/5} \\
           &\sim 10^{16} \left( \frac{B_{\text{p}}}{10^{8}\text{G}} \right)^{2/5} \left( \frac{\Omega}{0.1 \text{s}^{-1}} \right)^{4/5} \left( \frac{R}{10^{8.7}\text{cm}} \right)^{6/5} \left( \frac{t}{\text{yr}} \right)^{3/5} \text{cm}.
\end{split}
\end{equation}

The outer shock finally decays when the pressure of the shocked region $P_{\text{sh}}$ 
becomes equal to that of the ISM $p$. At this stage the 
shocked region may be physically continuous to the ISM. Solving Eq.(\ref{windEOM}) and (\ref{windEcon}) for $P_{\text{sh}}$,
\begin{equation}\label{P_s}
\begin{split}
P_{\text{sh}} &= \frac{7}{25} \left ( \frac{125}{154\pi} \right)^{2/5} \rho{}^{3/5} L^{2/5} t^{-4/5} \\
    &\sim 10^{-8} \left( \frac{B_{\text{p}}}{10^{8}\text{G}} \right)^{4/5} \left( \frac{\Omega}{0.1 \text{s}^{-1}} \right)^{8/5} \left( \frac{R}{10^{8.7}\text{cm}} \right)^{12/5} \left( \frac{t}{\text{yr}} \right)^{-4/5} \text{dyn/cm}^{-2}.
\end{split}
\end{equation}
Besides assuming that the density of the ISM is $\rho \sim 10^{-24} \text{g} \ \text{cm}^{-3}$, that is the number density of hydrogen 
is $n \sim 1 \text{cm}^{-3}$, the pressure can be estimated as 
\begin{equation}\label{p}
p = nk_{\text{B}} T \sim 10^{-13} \left( \frac{T}{10^3 \text{K}} \right) \text{dyn/cm}^{-2},
\end{equation}
where $k_{\text{B}} = 1.4 \times 10^{-16} \text{erg} \ \text{K}^{-1}$ is the Boltzmann constant and $T$ is the temperature of the ISM. 
From Eq.(\ref{P_s}) and (\ref{p}), the outer shock decays at about 
\begin{equation}\label{decay}
t_{\text{dec}} \sim 10^{6} \left( \frac{T}{10^3 \text{K}} \right)^{5/4} \text{yr},
\end{equation}
for the fiducial parameters of the WD pulsars. The lifetime of a pulsar $\tau$ can be estimated as 
\begin{equation}\label{tau}
\tau = \frac{E_{\text{rot}}}{L},
\end{equation} 
From Eq.(\ref{Erot2}) and (\ref{Lspindown}), for the fiducial parameters of the WD pulsars 
\begin{equation}\label{WDlifetime}
\tau_{\text{WD}} \sim 10^9 \left( \frac{M}{1.0 M_{\odot}} \right) \left( \frac{B_{\text{p}}}{10^8 \text{G}} \right)^{-2} \left( \frac{\Omega}{0.1\text{s}^{-1}} \right)^{-2} \left( \frac{R}{10^{8.7} \text{cm}} \right)^{-4} \text{yr}. 
\end{equation}
Compared with Eq.(\ref{decay}) and (\ref{WDlifetime}), we found that the outer shock decays at a very early stage of the lifetime of WD 
pulsars. 

For $t < t_{\text{dec}}$, the momentum transfer by the wind balances the pressure of shocked region at the inner shock front, 
\begin{equation}\label{Rs}
\frac{L c}{4\pi R_{\text{in}}{}^2}= P_{\text{sh}}.
\end{equation}
Then the radius of the inner shock front can be estimated as
\begin{equation}\label{Rin1}
\begin{split}
R_{\text{in}}(t < t_{\text{dec}}) &= \left( \frac{25}{28\pi} \right)^{1/2} \left( \frac{154\pi}{125} \right)^{1/5} \left( \frac{L}{\rho c^{5/3}} \right)^{3/10} t^{2/5} \\ 
                    &\sim 10^{15} \left( \frac{t}{\text{yr}} \right)^{2/5} \text{cm},
\end{split}
\end{equation}
for the fiducial parameters. For $t > t_{\text{dec}}$, there is no well-defined shocked region any more and the radius of the inner shock front is 
determined by the balance between the wind pressure and the pressure of the ISM $p$ instead of $P_{\text{sh}}$, and $R_{\text{in}}(t)$ become constant 
for $t$. For the fiducial parameters,
\begin{equation}\label{Rin_2}
R_{\text{in}}(t > t_{\text{dec}}) \sim 10^{17} \text{cm}.
\end{equation}
In the case of NS pulsars, the adiabatic cooling due to the expansion of the shocked region is considerable as a cooling process in the pulsar wind nebula. However, in the case of WD pulsars, since the outer edge of the shocked region 
does not expand after $t \gtrsim t_{\text{dec}}$, the adiabatic cooling shall give minor contributions to the cooling process of the high energy $e^{\pm}$.

Now we discuss the $e^{\pm}$ radiative cooling in the shocked region $r > R_{\text{in}}$. In the region swept by the shock, the magnetic field may be highly 
fluctuated and the high energy $e^{\pm}$ coming from the wind region are trapped because of the multiple scattering by the field, and lose the 
energy by the synchrotron radiation and inverse Compton scattering. Here we take the Bohm limit, where the fluctuation of the magnetic field 
$\delta B$ is comparable to the coherent magnetic field strength $B$. In this limit, the diffusion coefficient $D_{\text{sh}}$ can be approximated by 
\begin{equation}
D_{\text{sh}} = \frac{c r_{\text{g}}}{3},
\end{equation}
where $r_{\text{g}} = \epsilon/eB$ is the Larmor radius of the $e^{\pm}$ with energy $\epsilon$. The time scale $t_{\text{dif}}$ for 
the $e^{\pm}$ trapping in the shocked region is given by 
\begin{equation}\label{tdiff}
t_{\text{dif}} = \frac{d^2}{2D_{\text{sh}}} = \frac{3}{2} \frac{eBd^2}{\epsilon c},
\end{equation}
where $d$ is the size of the shocked region. 

We consider the age $t = \tau_{\text{WD}} > t_{\text{dec}}$. For $t > t_{\text{dec}}$, we set the size of the shocked region as the forward shock front at $t = t_{\text{dec}}$, 
that is 
\begin{equation}\label{D_WD}
d \approx R_{\text{out}}(t=t_{\text{dec}}) \sim 10^{19} \text{cm},
\end{equation}
for the fiducial parameters. As we have shown in Eq.(\ref{Rin_2}), the radius of the inner shock front is about $R_{\text{in}} \sim 10^{17} \text{cm}$ at 
$t = \tau_{\text{WD}}$. From Eq.(\ref{Bconf2}), the strength of the magnetic field at the inner edge $B_{\text{in}}$ can be estimated as 
\begin{equation}\label{B_in_WD}
B_{\text{in}} \sim 3 \times 10^{-6} \left( \frac{R_{\text{in}}}{10^{17}\text{cm}} \right)^{-1} \text{G},
\end{equation}
which is almost the same as that of the ISM. Then, substituting Eq.(\ref{D_WD}) and Eq.(\ref{B_in_WD}) to Eq.(\ref{tdiff})~\footnote{In this case, the diffusion coefficient can be estimated as 
\begin{equation}
D_{\text{sh}} \sim 10^{24} \left( \frac{\epsilon}{3 \mathrm{GeV}}\right) \mathrm{cm^2/s}.
\end{equation}
This $D_{\text{sh}}$ is smaller than the diffusion coefficient in the ISM~(see Eq.(\ref{diffusion})), which means that we consider the situation where the $e^{\pm}$ are highly trapped in the shocked region.}
, the time scale for the high energy $e^{\pm}$ with energy $\epsilon$ being trapped in the shocked region is
\begin{equation}
t_{\text{dif}} \sim 3 \times 10^{4} \left( \frac{\epsilon}{10\text{TeV}} \right)^{-1} \text{yr}.
\end{equation}
The synchrotron energy loss of the $e^{\pm}$ with energy $\epsilon$ is described as
\begin{equation}\label{synchrotron}
\frac{d \epsilon}{d t} = -\frac{4}{3}\sigma_{\text{T}} c \beta^2 \frac{B^2}{8\pi} \left( \frac{\epsilon}{m_{\text{e}} c^2} \right)^2,
\end{equation}
where $\sigma_{\text{T}}$ is the Thomson scattering cross section, and $\beta = v_{\text{e}}/c$ is the velocity in terms of the speed of light. Then from Eq.(\ref{synchrotron}), the typical energy loss of the electron with energy $\epsilon$ during the time scale $t_{\text{dif}}$ can be estimated as, 
\begin{equation}\label{Eloss}
\frac{\Delta \epsilon}{\epsilon} \sim 0.1 \left( \frac{B_{\text{in}}}{3 \times 10^{-6} \text{G}} \right)^3.
\end{equation}
This means that the high energy $e^{\pm}$ injected into the shocked region lose roughly $10 \%$ of the energy by the synchrotron radiation before diffusing out into the ISM. Although the inverse Compton scattering is also considerable process as a radiative cooling, it would be comparable to the synchrotron cooling. Then we can conclude that the radiative energy loss of $e^{\pm}$ in the pulsar wind nebula is not so large.

\subsection{Differences between white dwarf and neutron star pulsars}\label{Difference_in_NS_WD}
In this subsection, we discuss the differences between WD pulsars and NS pulsars as TeV $e^{\pm}$ sources.
 
Ordinary NS pulsars have been already discussed as a candidate for high energy $e^{\pm}$ sources
for the PAMELA positron excess~(\cite{Kawanaka:2009dk} and the 
references listed in Sec.\ref{sec1}). Compared with the NS pulsars, there are distinct features of the WD pulsars as high energy $e^{\pm}$ 
sources. As we saw in the previous sections, the WD pulsars can provide the high energy $e^{\pm}$ and the intrinsic energy budgets are almost the same as that of the NS pulsars. However, the magnetic field and rotation speed of the WD pulsars are much smaller than that of the NS pulsars. As a result, the spin down luminosity (Eq.(\ref{Lspindown})) of the WDs are much smaller than that of the NSs,
\begin{equation}
L_{\text{WD}} \sim 10^{41} \left(\frac{B_{\text{p}}}{10^8 \text{G}} \right)^2 \left(\frac{\Omega}{0.1 \text{s}^{-1}} \right)^{4} \left(\frac{R}{10^{8.7} \text{cm}} \right)^6 \text{erg/yr} \sim 10^{-4} L_{\text{NS}}.
\end{equation}
Then from Eq.(\ref{tau}), the lifetime of the WD pulsars are much longer than the NS pulsars 
\begin{equation}
\tau_{\text{WD}} \sim 10^9 \text{yr} \sim 10^4 \tau_{\text{NS}}. 
\end{equation}
Therefore, the number of the WD pulsars which are currently TeV $e^{\pm}$ sources are much larger than that of the NS pulsars. Since the high energy 
electrons above TeV cannot propagate more than $\sim 1$ kpc in our Galaxy, the number density of the WD pulsars which can be the TeV $e^{\pm}$ 
sources is 
\begin{equation}\label{nWD}
n_{\text{WD}} = \frac{\alpha \cdot \eta_{\text{WD}} \cdot \tau_{\text{WD}}}{V_{\text{G}}} \sim 10^3 \alpha \left( \frac{\eta_{\text{WD}}}{10^{-2} \text{yr}^{-1} \text{galaxy}^{-1}} \right) \left( \frac{\tau_{\text{WD}}}{10^{9} \text{yr}} \right) \left( \frac{V_{\text{G}}}{10^3 \text{kpc}^{3}} \right)^{-1} \text{kpc}^{-3}.
\end{equation}
where $V_{\text{G}}$ is the volume of our Galaxy and $\eta_{\text{WD}}$ is the event rate of the double degenerate WD binary merger in our Galaxy, 
Eq.(\ref{eta}). A parameter $\alpha$ is the fraction of the binary mergers which lead to the WD pulsars with the 
strong magnetic field $B \gtrsim 10^{8} \text{G}$. Eq.(\ref{nWD}) means 
that there may be enough WD pulsars which supply TeV $e^{\pm}$ near the Earth, 
although the parameter $\alpha$ has a large ambiguity at this stage. On the other hand, the number density 
of the TeV $e^{\pm}$ sources for the NS pulsars is 
\begin{equation}\label{rate_in_Galaxy}
n_{\text{NS}} \sim 0.1 \text{kpc}^{-3} \sim 10^{-4} \alpha^{-1} n_{\text{WD}}.
\end{equation} 
Eq.(\ref{rate_in_Galaxy}) means that it is 
uncertain whether NS pulsars are $e^{\pm}$ sources above TeV energy or not.

Another important difference is the environment of the pulsars, especially the strength of the magnetic field in the pulsar wind nebulae. The magnetic field is crucial for the cooling process since it determines how the high energy $e^{\pm}$ produced at pulsars are trapped and lose 
their energy by synchrotron radiation in the pulsar wind nebulae. In the case of the WD pulsars, the strength of the magnetic field at the 
shocked region is, in most of their lifetime, comparable to that of the ISM. As we saw in the previous subsection, this may imply that 
most of the accelerated $e^{\pm}$ directly escape into the ISM 
without cooling in the shocked region. On the other 
hand, in the NS pulsar wind nebulae, the situation is quite different. First the magnetic field are much stronger than the WD 
pulsars. Second there exists a SN shock front outside the pulsar wind nebula. These facts make the cooling process in the pulsar wind nebula more 
complicated, and the escape process of $e^{\pm}$ into the ISM is still uncertain. 

In the case of the NS pulsars, almost all the spin down luminosity is transformed to the kinetic energy of the $e^{\pm}$ wind before 
the wind goes into the shocked region~\cite{KennelCoroniti:1984}. The NS pulsars are consistent to be the source of the observed $e^{\pm}$ if 
the $e^{\pm}$ lose $\sim 99 \%$ of their energy in the shocked region~\cite{Kawanaka:2009dk}. As we discussed in Sec.\ref{sec2}, 
the total energy budgets of the WD and NS pulsars are almost the same when almost all the double degenerate WD binaries 
merge to become the WD pulsars, that is when $\alpha = 1$. Since the $e^{\pm}$ lose only $\sim 10 \%$ of their energy in the WD pulsar 
wind nebulae~(Eq.(\ref{Eloss})), the expected amount of $e^{\pm}$ from WD pulsars can exceed the current observation bound. Hence if 
\begin{equation}\label{fiducial_alpha}
\alpha \sim 0.01, 
\end{equation}
we can expect that the PAMELA positron excess can be explained by the WD pulsars without any contribution of other sources. We should note 
that the fraction in Eq.(\ref{fiducial_alpha}) seems consistent with the observed fraction of the magnetized WDs $\sim 10 \%$. 
We show the brief summary of the comparison between WD and NS pulsars in Table \ref{comparison}.

\begin{table}[h]
\begin{center}
{\renewcommand\arraystretch{1.5}
\begin{tabular}{c|c|c|c|c|c|c|c}
 & \shortstack{energy per each \\ $E_{\text{rot}} \ \mathrm{[erg]}$} &  \shortstack{luminosity per each \\ $L \ \mathrm{[erg/yr]}$} & \shortstack{lifetime \\ $\tau \ \mathrm{[yr]}$} & \shortstack{event rate \\ $\mathrm{[1/yr/galaxy]}$} & \shortstack{number density \\ $n \ [1/\mathrm{kpc}^{3}]$} & \shortstack{efficiency \\ $\mathrm{[\%]}$} & \shortstack{$L \times n \times \mathrm{efficiency}$ \\ $\mathrm{[erg/yr/kpc^3]}$}   \\ \hline 
WD pulsar & $\sim 10^{50}$ & $\sim 10^{41}$ & $\sim 10^{9}$ & $\sim \alpha/100$ & $\sim 10^{3}\alpha$ & $\sim 90$ & $\sim 10^{44} \alpha$ \\  
NS pulsar & $\sim 10^{50}$ & $\sim 10^{45}$ & $\sim 10^{5}$ & $\sim 1/100$ & $\sim 0.1$ & $\sim 1$ & $\sim 10^{42}$ \\  \hline
\end{tabular}
}
\caption{The comparison between WD and NS pulsars as $e^{\pm}$ sources.}
\label{comparison}
\end{center}
\end{table}

\section{Energy Spectrum Calculation}\label{sec3}
In this section, we calculate the $e^{\pm}$ energy spectrum observed at the solar system after the propagation in our Galaxy 
for the WD pulsar model. We solve the diffusion equation taking into account the Klein-Nishina~(KN) effect.

\subsection{electron distribution function from a single source}
Here we formulate cosmic ray $e^{\pm}$ propagation through our Galaxy according to \cite{Atoyan:1995}. For simplicity we assume that 
the diffusion approximation is good (e.g., neglecting convection), the $e^{\pm}$ propagate in spherically symmetric way and diffuse homogeneously in our Galaxy~\footnote{These assumptions become worse as the energy of $e^{\pm}$ decreases below $\lesssim 10 \mathrm{GeV}$. We discuss the validity of our results by comparing with more realistic calculation using the GALPROP code~\cite{GALPROP} in Sec.\ref{sec4}.}. Following these assumptions, the $e^{\pm}$ propagation equation can be described as follow.
\begin{equation}\label{diff_eq}
\frac{\partial f}{\partial t}=\frac{D}{r}\frac{\partial }{\partial r}r^2\frac{\partial f}{\partial r}- \frac{\partial }{\partial \epsilon}(Pf)+Q.
\end{equation}
Here $f(t,\epsilon,r)~\mathrm{[m^{-3} \cdot GeV^{-1}]}$ is the energy distribution function of $e^{\pm}$. $P(\epsilon)$ is the cooling function of the $e^{\pm}$ 
which corresponds to the energy loss rate during the propagation. $D(\epsilon)$ denotes the diffusion coefficient, which does not depend on the 
position $r$. $Q(t,\epsilon,r)$ is the energy injection term. Considering $\delta$-function injection at the time $t=t_0$, that is 
\begin{equation}
Q(t,\epsilon,r) = \Delta N(\epsilon) \delta(r)\delta(t-t_0),
\end{equation}
we can get the analytical solution~\cite{Atoyan:1995}. For an arbitrary injection spectrum $\Delta N(\epsilon)$, the energy distribution 
can be described as 
\begin{equation}\label{dis_short}
f(r,t,\epsilon) = \frac{\Delta N(\epsilon_{t,0})}{\pi^{3/2}r_{\text{dif}}{}^3}\frac{ P(\epsilon_{t,0})}{P(\epsilon)}\exp \left( -\frac{r^2}{r_{\text{dif}}{}^2} \right).
\end{equation}
Here $\epsilon_{t,0}$ corresponds to the energy of $e^{\pm}$ which are cooled down to $\epsilon$ during the time $t-t_0$, and is obtained
by solving the integral equation
\begin{equation}
t-t_0= \int^{\epsilon_{t,0}}_{\epsilon} \frac{d \epsilon'}{P(\epsilon')}.
\end{equation}
The $e^{\pm}$ propagate to the diffusion length defined by
\begin{equation}\label{r_diff}
r_{\text{dif}}(\epsilon, \epsilon_{t,0}) = 2\left( \int^{\epsilon_{t,0}}_{\epsilon}
\frac{D(\epsilon')}{P(\epsilon')} d \epsilon' \right)^{1/2}.
\end{equation}
Eq.(\ref{dis_short}) is the distribution function for the $\delta$-functional (short term) injection source, i.e., the Green function of 
Eq.(\ref{diff_eq}). From now on, we set the observation is taking place at $t=0$. 

Even for a continuous (long term) injection source, the distribution function can be calculated by 
integrating Eq.(\ref{dis_short}) for the active time of the source. The integration can be done numerically by
transforming the integration from $dt_0$ to $d \epsilon_{t,0} = P(\epsilon_{t,0}) dt_0$. (That is, we take $\epsilon_{t,0}$ as the time coordinate.) 
Substituting $\Delta N(\epsilon_{t,0}(\epsilon,t_0)) = Q(\epsilon_{t,0}(\epsilon,t_0))dt_0$ into Eq.(\ref{dis_short}) and integrating over $dt_0$, 
the resulting distribution function reads
\begin{equation}\label{dis_long}
f(\epsilon,r) = \frac{1}{\pi^{3/2} P(\epsilon)}\int^{\epsilon_{\hat{t}}}_{\epsilon}\frac{Q(\epsilon_{t,0})}{r_{\text{dif}}(\epsilon,\epsilon_{t,0})^3}\exp \left( -\frac{r^2}{r_{\text{dif}}(\epsilon,\epsilon_t)^{2}} \right)d \epsilon_{t,0}.
\end{equation}
Here $\epsilon_{\hat{t}}$ is the energy of $e^{\pm}$ when they leave the source at the source birth time $t=\hat{t}~(<0)$, that is
\begin{equation}
\hat{t} = -\int^{\epsilon_{\hat{t}}}_{\epsilon} \frac{d \epsilon'}{P(\epsilon')}.
\end{equation}
The flux at $r$ is given by $\Phi(\epsilon,r)=(c/4\pi) f(\epsilon,r) \mathrm{[m^{-2} \cdot s^{-1} \cdot sr^{-1} \cdot GeV^{-1}]}$. 

Now, in order to estimate the observed $e^{\pm}$ flux, we have to specify the 
cooling function $P(\epsilon)$, diffusion coefficient $D(\epsilon)$ and injected energy spectrum $Q(\epsilon)$. 
First, we formulate the $e^{\pm}$ cooling function including the KN effect.
Following the equation (5) in \cite{Stawarz:2009ig}, the energy loss rate of the $e^{\pm}$ including the KN effect is written as 
\begin{equation}\label{cooling_KN}
P(\epsilon) = -\frac{d \epsilon}{dt}= \frac{4}{3}\sigma_{\text{T}} c \left( \frac{\epsilon}{m_{\text{e}} c^2} \right)^2 \left[ \frac{B_{\text{ISM}}^2}{8\pi} + \int d \epsilon_{\text{ph}} u_{\text{tot}}(\epsilon_{\text{ph}}) f_{\text{KN}}\left(\frac{4 \epsilon \epsilon_{\text{ph}}}{m_{\text{e}}{}^2 c^4} \right) \right].
\end{equation}
Here $\sigma_{\text{T}} = 6.62 \times 10^{-25} \mathrm{cm^2}$ is the Thomson scattering cross section and 
$\epsilon_{\text{ph}}$ are the energy of the background photon. 
$B_{\text{ISM}}$ is the magnetic field strength in the ISM
where we set $B_{\text{ISM}} = 1 \mathrm{\mu G}$.
$f_{\text{KN}}$ is the KN suppression function which is explicitly shown in \cite{Moderski:2005}.
\begin{equation}
f_{\text{KN}}(\tilde{b}) = \frac{9g(\tilde{b})}{\tilde{b}^3},
\end{equation}
where
\begin{equation}
g(\tilde{b}) = \left( \frac{1}{2}\tilde{b} +6 +\frac{6}{\tilde{b}}  \right)\ln (1+\tilde{b})-\left(\frac{11}{12}\tilde{b}^3 +6\tilde{b}^2+ 9\tilde{b}+4 \right) \frac{1}{(1+\tilde{b})^2} - 2 + 2\text{Li}_2(-\tilde{b})
\end{equation}
and $\text{Li}_2$ is the dilogarithm
\begin{equation}
\text{Li}_2(z) = \int^{0}_{z}\frac{\ln(1-s)ds}{s}.
\end{equation}
The ISM photons consists of the stellar radiation, reemitted radiation from dust, and CMB,
\begin{equation}
u_{\text{tot}} = u_{\text{star}} + u_{\text{dust}} + u_{\text{CMB}}.
\end{equation} 
Here we model the interstellar radiation field using the results of the GALPROP code~\cite{Poter:2008}. 
\begin{figure}[htbp]
 \begin{center}
  \includegraphics[width=70mm]{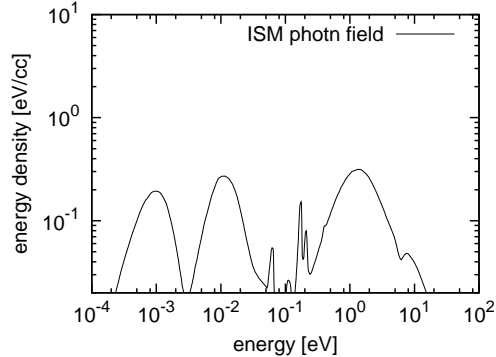}
 \end{center}
 \caption{The energy density of the ISM photon field at $8\mathrm{kpc}$ from the center of our Galaxy~\cite{Poter:2008}.}
 \label{fig:photon}
\end{figure}
Fig.\ref{fig:photon} shows the ISM radiation field energy density $\epsilon_{\text{ph}} \times u_{\text{tot}}(\epsilon_{\text{ph}})$ at $\sim 8\mathrm{kpc}$ from the center of our Galaxy. Following the formulation above, we numerically calculate the $e^{\pm}$ cooling function including the KN effect.
\begin{figure}[htbp]
 \begin{center}
  \includegraphics[width=70mm]{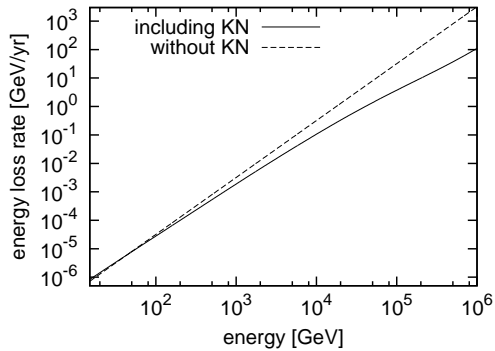}
 \end{center}
 \caption{The cooling function $P(\epsilon)$ for $e^{\pm}$ at $\sim 8$ kpc from the center of our Galaxy.}
 \label{fig:cooling}
\end{figure}
Fig.\ref{fig:cooling} shows the cooling function for $e^{\pm}$ with or without the KN effect. The solid line shows the function $P(\epsilon)$ 
in Eq.(\ref{cooling_KN}). The dotted line shows the cooling function when we set $f_{\text{KN}} = 1$. We can see that the KN effect becomes relevant for 
$\epsilon \gtrsim 1 \mathrm{TeV}$.

Second, we formulate the diffusion coefficient. As the diffusion coefficient $D(\epsilon)$ for the $e^{\pm}$ propagating through our Galaxy, 
we use an empirical law given by the boron-to-carbon ratio observation, that is
\begin{equation}\label{diffusion}
D(\epsilon)=D_0\left(1+ \frac{\epsilon}{3\mathrm{GeV}} \right)^{\delta}.
\end{equation}
Here $D_0 = 5.8 \times 10^{28} \mathrm{cm^2 s^{-1}} $, $\delta=1/3$~\cite{Baltz:1998xv}.

Finally, we assume that the intrinsic energy spectrum at the source is described by the cutoff power law, that is
\begin{equation}\label{injection}
Q(\epsilon,t_0,\hat{t}) = Q_0 \epsilon^{-\nu} \exp \left( -\frac{\epsilon}{\epsilon_{\text{cut}}} \right) \left( 1+\frac{t_0-\hat{t}}{\tau} \right)^{-2}.
\end{equation}
Here $\tau$ is the lifetime of the source and $t_0$ is the time when the $e^{\pm}$ leave the source. Then substituting Eq.(\ref{cooling_KN}), 
(\ref{diffusion}) and (\ref{injection}) into Eq.(\ref{dis_long}), we can get the observed electron distribution function $f(\epsilon,r,\hat{t})$ from 
a pulsar which is located at the distance $r$ from the solar system and born at $t=\hat{t}$.

\subsection{$e^{\pm}$ distribution function from multiple sources}
Here we consider the $e^{\pm}$ distribution function from multiple sources. As we show in Sec.\ref{Difference_in_NS_WD}, there should be multiple pulsars which contribute to the observed $e^{\pm}$ flux. 

To calculate the distribution function from multiple sources, 
we integrate Eq.(\ref{dis_long}) for the pulsar birth time $\hat{t}$ 
and the pulsar position $r$, taking into account the birth rate of the pulsars. Then the observed $e^{\pm}$ distribution function is
\begin{equation}\label{multi_spec}
F(\epsilon) = \int^{0}_{-\tau_{\text{WD}}} d \hat{t} \int^{r_{\text{dif}}(\epsilon,\epsilon_{\hat{t}})}_{0} 2 \pi r dr \cdot \alpha \cdot \eta_{\text{WD}} f(\epsilon, r, \hat{t}).
\end{equation}
Again $\eta_{\text{WD}}$ is the merger rate of the double degenerate WD binary, and $\alpha$ is the fraction 
of the mergers resulting in WD pulsars. We take the lifetime of the WD pulsars $\hat{t}=-\tau_{\text{WD}}$ as the lower limit of the time integral. 
We have comfirmed that the following results does not depends on this limit as long as it is smaller than $-\tau_{\text{WD}}$.
As the upper limit of the space integral we take the diffusion length $r_{\text{dif}}(\epsilon ,\epsilon_{\hat{t}})$, which is defined 
in the same way as Eq.(\ref{r_diff}). Through the distance $r_{\text{dif}}(\epsilon, \epsilon_{\hat{t}})$, the energy of the propagating $e^{\pm}$ 
changes from $\epsilon_{\hat{t}}$ to $\epsilon$. 

Since Eq.(\ref{multi_spec}) is the mean value,
we also estimate the standard deviation 
of the calculated energy spectrum, that is 
\begin{equation}\label{delta_F}
(\delta F)^2 = \int^{0}_{-\tau_{\text{WD}}} d \hat{t} \int^{r_{\text{dif}}(\epsilon ,\epsilon_{\hat{t}})}_{0} 2 \pi r dr \cdot \alpha \cdot \eta_{\text{WD}} f^2 - N f_{\text{ave}}^2 ,
\end{equation}
where $N$ is the number of the pulsars in our Galaxy that are the source of observing $e^{\pm}$, that is 
\begin{equation}
N = \int^{0}_{-\tau_{\text{WD}}} d \hat{t} \int^{r_{\text{dif}}(\epsilon ,\epsilon_{\hat{t}})}_{0} 2 \pi r dr \cdot \alpha \cdot \eta_{\text{WD}} ,
\end{equation}
and $f_{\text{ave}} = F(\epsilon)/N$ is the averaged $e^{\pm}$ spectrum per pulsar.
We should note that the integral of Eq.(\ref{delta_F}) contains a serious 
divergence at $\hat{t}=0$ because of the large but improbable contributions from very young and nearby sources~\cite{lee79,Berezinski:1990,lagutin95,Ptuskin:2006}. Here we follow Ptuskin et al (2006) and set the cutoff parameter as
\begin{equation}
\hat{t}_{\text{c}}(\epsilon) = -(4\pi \eta_{\text{WD}} \cdot \alpha D(\epsilon) )^{-1/2},
\end{equation}
which approximately corresponds to the birth time of the newest pulsr that contributes $e^{\pm}$ with energy $\epsilon$.


\subsection{Results}

\begin{figure}[htbp]
\centering
\subfigure[The $e^{\pm}$ total flux]{\includegraphics[width=110mm]{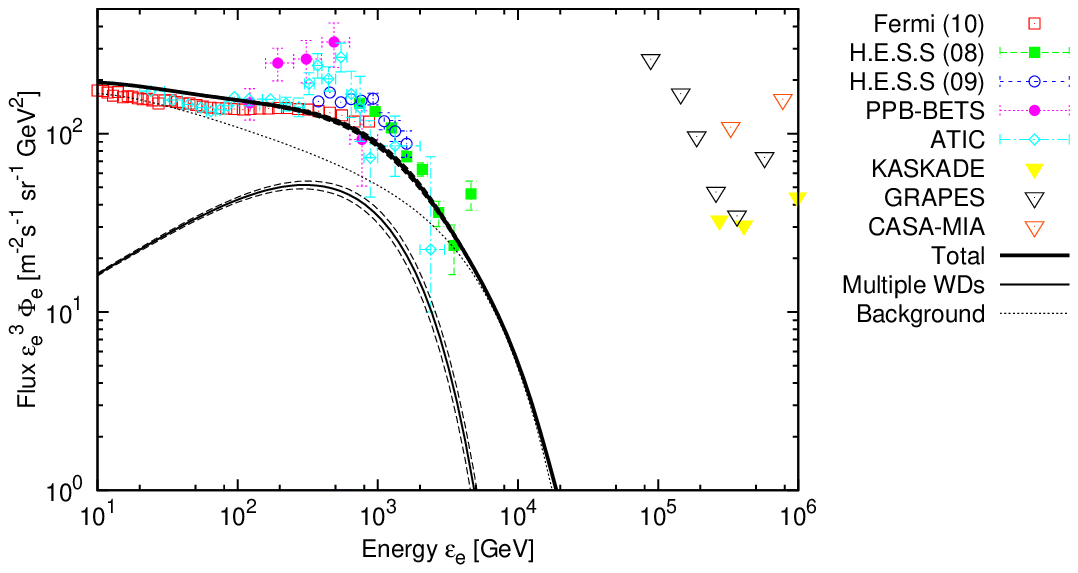}}
\subfigure[The positron fraction]{\includegraphics[width=60mm]{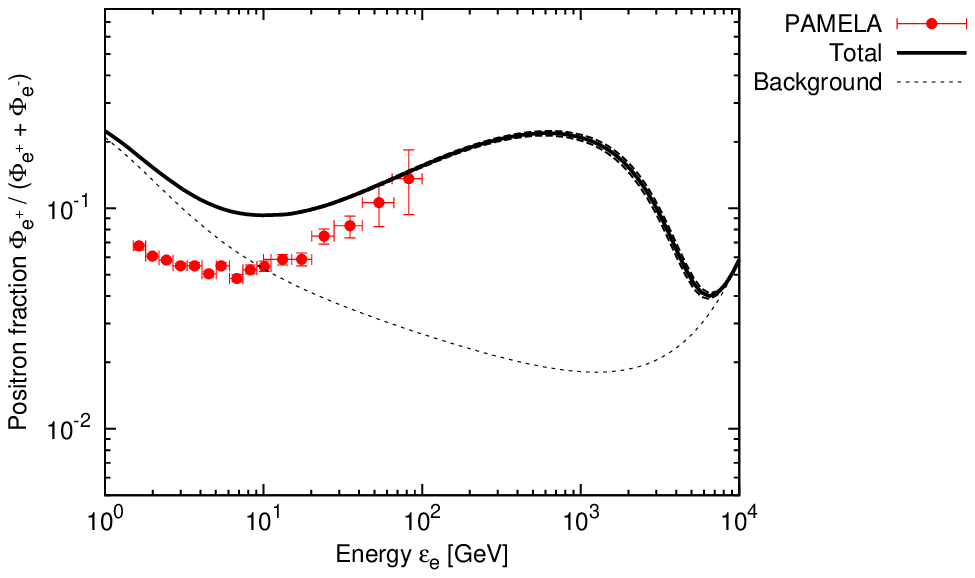}}
\caption{WD pulsar dominant model : The left panel shows the energy spectrum of the $e^{\pm}$ including the contributions of WD pulsars~(thick solid line). In this model, we set the cutoff energy of the injection spectrum for each WD pulsar $\epsilon_{\text{cut}} \sim 1 \mathrm{TeV}$. The average flux (thin solid line), flux with the standard deviation (thin dashed lines) and background (dotted line) are shown. For the background flux, we adopt the fitting function in Baltz \& Edsj$\ddot{\text{o}}$ (1999)~\cite{Baltz:1998xv} with an exponential cutoff for the primary electron flux at $5 \mathrm{TeV}$, which is similar to
that shown in Aharonian et al. (2008)~\cite{Collaboration:2008aa}. We assume that each WD pulsar emits the same amount of $e^{\pm}$. We set the total energy $\sim 10^{50} \mathrm{erg}$ for each WD pulsar, intrinsic spectral index $\nu = 1.9$, lifetime of WD pulsars $\sim 10^{9} \mathrm{yr}$ and birth rate in our Galaxy $\alpha \cdot \eta_{\text{WD}} \sim 10^{-7} \mathrm{yr}^{-1} \mathrm{kpc}^{-2}$. The right panel shows the positron fraction resulting from the average spectrum (solid line) with the dispersion (dushed lines) and background (dotted line), compared with the PAMELA data. The background contribution begins to rise around $\sim 3\mathrm{TeV}$ since we set the exponential cutoff only for the primary electron background, not for the secondary $e^{\pm}$ background. Note also that the solar modulation is important below $\sim 10 \mathrm{GeV}$.}\label{model_A}
\vspace{4mm}
\end{figure}

\begin{figure}[htbp]
\centering
\subfigure[The $e^{\pm}$ total flux]{\includegraphics[width=110mm]{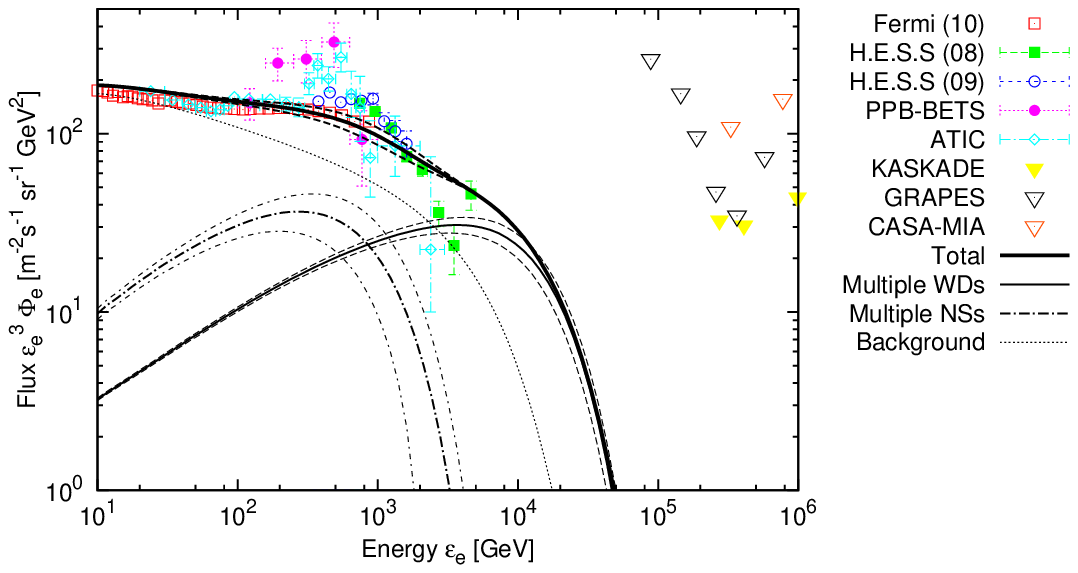}}
\subfigure[The positron fraction]{\includegraphics[width=60mm]{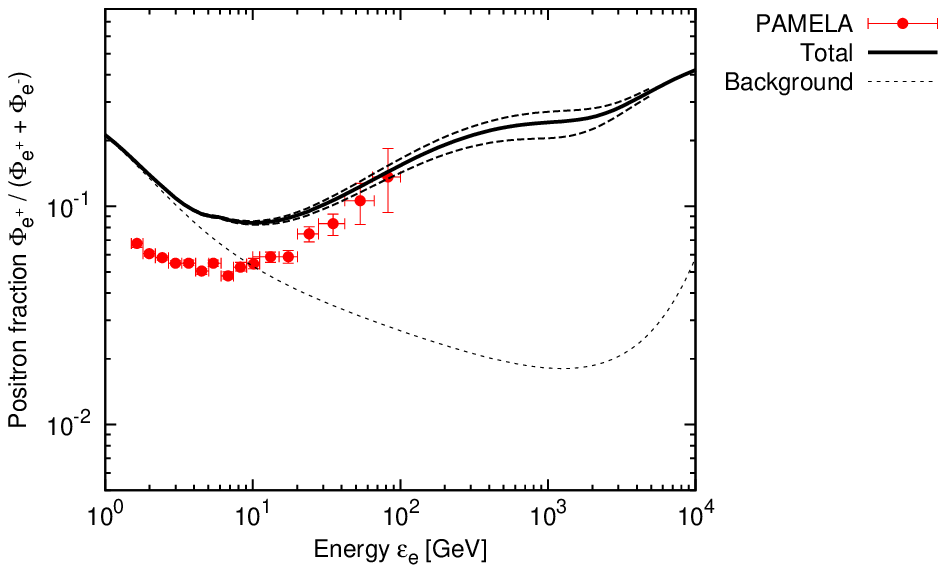}}
\caption{WD and NS pulsar mixed model : The left panel shows the energy spectrum of the $e^{\pm}$~(thick solid line) including the contributions of WD pulsars~(thin solid line) with the same parameters as Fig.\ref{model_A} except for the total energy for each $\sim 5 \times 10^{49} \mathrm{erg}$ and cutoff energy of the injection spectrum $\epsilon_{\text{cut}} \sim 10 \mathrm{TeV}$. The contribution of multiple NS pulsars~(dotted dash line) with the total energy $\sim 10^{48} \mathrm{erg}$ for each NS, cutoff energy of the injection $\epsilon_{\text{cut}} \sim 1 \mathrm{TeV}$, lifetime $\sim 10^{5} \mathrm{yr}$, birth rate in our Galaxy $\sim 10^{-5} \mathrm{yr}^{-1} \mathrm{kpc}^{-2}$ are also included. For both fluxes of pulsars, the standard deviations are shown. The total $e^{\pm}$ flux and its deviation are the thick solid and dushed line, respectively. The right panel shows the positron fraction.}  
\label{model_B}
\end{figure}

Here we consider two types of models. Fig.\ref{model_A} shows the WD pulsar dominant model. In the left panel, the $e^{\pm}$ flux from multiple WD pulsars~(thin solid line) is shown with the standard deviations~(thin dashed line), background flux~(dotted line) and total flux~(thick solid line). For each WD pulsar, we set the cutoff energy of the injection spectrum $\epsilon_{\text{cut}} \sim 1 \mathrm{TeV}$~(Eq.(\ref{injection})), intrinsic spectral index $\nu = 1.9$, lifetime $\tau_{\text{WD}} \sim 10^{9} \mathrm{yr}$, total energy for each $\sim 10^{50} \mathrm{erg}$, merger rate of double degenerate WD pulsar binaries $\eta = 10^{-5} \mathrm{yr}^{-1} \mathrm{kpc}^{-2}$ and probability of forming WD pulsars $\alpha = 0.01$, which means that the birth rate of WD pulsars in our Galaxy is $\sim 10^{-7} \mathrm{yr}^{-1} \mathrm{kpc}^{-2}$. The left panel of Fig.\ref{model_A} includes the observational data of cosmic ray electrons plus positrons given by the balloon and satellite experiments, ATIC/PPT-BETS/Fermi~\cite{chang:2008,Torii:2008xu,Ackermann:2010_8,Abdo:2009zk,Moiseev:2007js}, and also the data of ground-based air Cherenkov telescopes, H.E.S.S/KASKADE/GRAPES/CASA-MIA~\cite{Collaboration:2008aa,Aharonian:2009ah,Schatz:2003,Gupta:2009,Chantell:1997}. For KASKADE/GRAPES/CASA-MIA, the plots show the observed flux of the diffuse gamma rays. Since a gamma-ray entering into the air first produces an $e^{\pm}$ pair to begin a cascade, its shower will look very similar to that of an $e^{\pm}$ of equivalent energy~\cite{Kistler:2009wm}. Thus we presume these date as the upper limits on the $e^{\pm}$ flux. H.E.S.S electron data are also partly contaminated with photons. Therefore, a viable model should not significantly overshoot the points. The background flux consists of the primary electrons which is conventionally attributed to the SNRs and the secondary $e^{\pm}$ produced by the hadron interaction between cosmic ray protons and the interstellar matter, and successive pion decays. For the secondary $e^{\pm}$ flux, we adopt the fitting function in Baltz \& Edsj$\ddot{\text{o}}$~(1999)~\cite{Baltz:1998xv,Ptuskin:2006b,Moskalenko:1997gh}. For the primary electron flux, we also refer Baltz \& Edsj$\ddot{\text{o}}$~(1999) but with an exponential cutoff at $5 \mathrm{TeV}$~\footnote{We also reduce the flux by $30 \%$ since the fitting function of Baltz \& Edsj$\ddot{\text{o}}$~(1999) provide larger flux than the data even without other contributions.}, which is similar to that shown in Aharonian et al. (2008)~\cite{Collaboration:2008aa}. Our result fits well the observational data of H.E.S.S and Fermi.

The right panel of Fig.\ref{model_A} shows the positron fraction using the same parameters as the left panel. The results show that the observed 
positron excess can be explained by considering only the contribution from multiple WD pulsars, and the positron fraction is expected to drop at 
around the WD pulsar cutoff energy $\sim \mathrm{TeV}$. The background contribution of the positron fraction begins to rise around 
$\sim 3 \mathrm{TeV}$ since we set the exponential cutoff only for the primary electron background, not for the secondary $e^{\pm}$ background. This 
treatment is appropriate since the abundance of cosmic ray protons is observationally robust in this energy range and so is the amount of the 
secondary $e^{\pm}$ background. As we discuss in Sec.\ref{sec4}, our calculations become less reliable below $\lesssim 10 \mathrm{GeV}$ since we neglect the anisotropic effects during the diffusion in the Galactic disk. Note that in these energy range, the solar modulation is also relevant.

Fig.\ref{model_B} shows the WD and NS pulsar mixed model. In the left panel, the thin solid line shows the $e^{\pm}$ flux from multiple WD pulsars which have the total energy for each $\sim 5 \times 10^{49} \mathrm{erg}$, cutoff energy of the injection spectrum $\epsilon_{\text{cut}} \sim 10 \mathrm{TeV}$ and the same value for other parameters as Fig.\ref{model_A}. The difference of the $\epsilon_{\text{cut}}$ means, in our WD pulsar model, the difference of the multiplicity ${\cal M}$, the magnetic field strength $B_{\text{p}}$, the angular freauency $\Omega$ and the radius $R$ according to Eq.(\ref{emax}). The dotted-dash line shows the $e^{\pm}$ flux from multiple NS pulsars with the total energy $\sim 10^{48} \mathrm{erg}$ for each, cutoff energy of the injection energy spectrum $\epsilon_{\text{cut}} \sim 1 \mathrm{TeV}$, lifetime $\sim 10^{5} \mathrm{yr}$, birth rate in our Galaxy $\sim 10^{-5} \mathrm{yr}^{-1} \mathrm{kpc}^{-2}$. The standard deviation of the $e^{\pm}$ energy flux from the WD pulsars is relatively small compared with the NS pulsars. This is because the larger abundance of WD puslar are expected as we discussed in the previous sections. The dotted line shows the same background contribution as Fig.\ref{model_A}. The total flux and the deviations are shown by thick solid line and thick dushed line, respectively. It is shown that the excess in the $e^{\pm}$ flux in the range $100 \mathrm{GeV} \lesssim \epsilon \lesssim 1 \mathrm{TeV}$ is explained by the multiple NS pulsars.
 By considering the contribution of multiple WD pulsars, the smooth "double bump" are formed in the energy spectrum around $1 \mathrm{TeV}$ and $10 \mathrm{TeV}$, which can be observable by the future experiments like CALET \cite{torii:2006,torii:2008} and CTA~\cite{CTA:2010}. 

The right panel of Fig.\ref{model_B} shows the positron fraction for the mixed model. The observed positron excess can be explained and, in this case, there will be no flux drop around $\sim \mathrm{TeV}$ in contrast to Fig.~\ref{model_A}.

\subsection{Observed WD pulsar candidates}\label{observation}
Finally, in this subsection we give two interesting examples of the observed WD pulsar candidates, AE Aquarii and EUVE J0317-855. 
So far, a few thousand WDs have been discovered, and the magnetic field and the rotational period have been detected for some of them~\cite{Wickramasinghe:2000,Schmidt:2003,Vanlandingham:2005,Mereghetti:2009}. Forthcoming experiments like ASTRO-H~\cite{Takahashi:2008} will find more magnetized and rapidly spinning WDs, which will reveal the detailed characteristics of such WDs. Then, we will know whether sufficient amount of WD pulsars exist in our Galaxy or not. \\

\noindent \textbf{AE Aquarii} 

AE Aquarii is a magnetized cataclysmic variable, that is, consisting of a primary WD and a spectral type K5V main sequence star, 
located at $\sim 100 \mathrm{pc}$ from the solar system. The primary WD has spin period $\sim 33\mathrm{s}$, which is 
identified by the approximately sinusoidal profiles of the observed emissions at energies below 
$\sim 4 \mathrm{keV}$~\cite{Patterson:1979,Eracleous:1994}. Recently Suzaku satellite discovered that AE Aquarii shows hard X-ray 
sharp pulsations at the period consistent with its rotation~\cite{Terada:2007br}. Also TeV gamma emissions during the optical flares were reported~\cite{Meintjes:1992,Meintjes:1994}, although there have been no detection since then. The primary WD is spinning down at a rate 
$\sim 6 \times 10^{-14} \mathrm{sec} \ \mathrm{sec}^{-1}$, implying 
the spin down luminosity $\sim 10^{33} \mathrm{erg/sec}$, which is three orders of 
magnitude larger than the UV to X-ray emissions. The magnetic field strength inferred from the spin down luminosity is 
$\sim 5 \times 10^{7}\mathrm{G}$~\cite{Ikhsabov:1998}. 

Since the AE Aquarii is an accreting binary system, the density of the plasma surrounding the primary WD may be 
much higher than the GJ density. However, both theoretical~\cite{Wynn:1997} and observational works suggest that the rapid rotation and strong 
magnetic field produce a low-density region around the WD, and 
the particle acceleration by the same mechanism as spin-powered pulsars could be 
possible. The parameters of AE Aquarii satisfy the condition Eq.(\ref{avalanche_polar}), above the death line of WD pulsars 
(Fig.\ref{death_line}). \\

\noindent \textbf{EUVE J0317-855~(RE J0317-853)}

EUVE J0317-855 is a hydrogen-rich magnetized WD discovered by ROSAT and EUVE survey~\cite{Barstow:1995,Ferrario:1997}. 
By analyzing the photometric, spectroscopic and polarimetric variations, EUVE J0317-855 is shown to rotate at the period 
$\sim 725 \mathrm{s}$, which is one of the fastest isolated WDs, and the dipole magnetic field is $\sim 4.5 \times 10^{8} \mathrm{G}$. 
EUVE J0317-855 have a DA WD companion which is located at $\gtrsim 10^3 \mathrm{AU}$ from EUVE J0317-855. Because of the large separation, 
there suppose to be no interaction between the two WDs. By analysing the emission from the companion, Barstow et al (1995)~\cite{Barstow:1995} 
noted that EUVE J0317-855 is located at $\sim 35\mathrm{pc}$ from the solar system, and the mass is $1.31\mbox{-}1.37 M_{\odot}$ which is relatively 
large compared with the typical WD mass $\sim 0.6M_{\odot}$. Its rapid rotation and large mass suggest that EUVE J0317-855 may be the outcome of a 
double degenerate WD binary merger~\cite{Ferrario:1997}. Relevant pulse emission from EUVE J0317-855 has not been observed yet, which may 
suggest that the $e^{\pm}$ creation and acceleration does not occur. When we put the parameters of EUVE J0317-855 on Fig.\ref{death_line}, it comes below the death line, which is also consistent with the observation.

\section{Summary and Discussion}\label{sec4} 
We have investigated the possibility that WD pulsars become a new TeV $e^{\pm}$ source. We have supposed that a fair fraction of double degenerate WD binaries merge to become WD pulsars, and these WDs have the magnetospheres and pulsar wind nebulae. The $e^{\pm}$ pair creation in the magnetospheres and their acceleration and cooling in the wind nebulae have been discussed, and we have found the following.
\begin{enumerate}
\item If a double degenerate WD binary merges into a maximally spinning WD, its rotational energy will become $\sim 10^{50} \mathrm{erg}$, which is comparable to that of a NS pulsar. Also the birth rate $\sim 10^{-2} \mbox{-} 10^{-3} \mathrm{/yr/galaxy}$ is similar to the NS case, which provides the right energy budget for cosmic ray $e^{\pm}$.
\item Applying the theory of NS magnetospheres, we give the $e^{\pm}$ pair creation condition~("the death line") for WD pulsars. Since our fiducial parameters of WD pulsars meet the condition, the WD pulsars are eligible for the $e^{\pm}$ factories. The death line is consistent with the observations for some WD pulsar candidates.   
\item By assuming the energy equipartition between $e^{\pm}$ and magnetic field in the wind region, we have shown that the $e^{\pm}$ produced in the WD pulsar magnetosphere can accelerate up to $\sim 10 \mathrm{TeV}$ when the WD pulsar has a rapid rotation~($P \sim 50 \mathrm{s}$) and strong magnetic fields~($B \sim 10^{8} \mathrm{G}$) and the $e^{\pm}$ multiplicity is not so large~(${\cal M} \sim 1$).  
\item In contrast to the NS case, the adiabatic energy losses of $e^{\pm}$ in the pulsar wind nebula region are negligible in the case of the WD pulsars since they continue to inject the $e^{\pm}$ after the nebula stop expanding. Also the radiative cooling of $e^{\pm}$ is not so large, and the high energy $e^{\pm}$ can escape from the nebula without losing much energy. As a consequence, it is enough that a fraction $\sim 1 \%$ of WDs are magnetized as observed in order for the WD pulsars to become the relevant TeV $e^{\pm}$ sources.    
\end{enumerate}

Based on the WD pulsar model above, we have calculated the observed $e^{\pm}$ flux from multiple WD pulsars in our Galaxy. We have solved the diffusion equation including the KN effect, and found the following. 
\begin{enumerate}
\item[5.] We have shown the two model $e^{\pm}$ fluxes. In one model~(WD pulsar dominant model), only considering the contribution from the multiple WD pulsars, we can explain the reported excess of the $e^{\pm}$ flux around $100 \mathrm{GeV} \lesssim \epsilon \lesssim 1 \mathrm{TeV}$ and also the PAMELA positron excess. In the other model~(WD and NS pulsar mixed model), the combination of the multiple WD and NS pulsars can also explain the existent observations, and form the double bump in the energy spectrum of $e^{\pm}$, which can be a signature for the future $e^{\pm}$ observation like CALET \cite{torii:2006,torii:2008} and CTA~\cite{CTA:2010}. Since the lifetime of WD pulsars are relatively large, the number of nearby active sources can be huge, which give a small Poisson fluctuation in the $e^{\pm}$ flux compared with NS pulsars. 
\end{enumerate}

As we have shown, WD pulsars could dominate the quickly cooling $e^{\pm}$ above TeV energy as a second spectral bump or even surpass the NS pulsars in the observing energy range $\sim 100$ GeV, providing a background for the dark matter signals and a nice target fo the future AMS-02~\cite{Beischer:2009,Pato:2010ih}, CALET~\cite{torii:2006,torii:2008} and CTA~\cite{CTA:2010}. As the future works we should consider other observational signatures than $e^{\pm}$ for the coming multi-messenger astronomy era. For example we have to consider the radio to $\gamma$ ray emission from WD pulsars based on our model. The number of observed pulsars in the Galactic disk should be proportional to $\sim$ (number density) $\times$ (radio luminosity). Since about $\sim 10^3$ NS pulsars have been discovered by radio telescopes, assuming that WD pulsars can convert the spin down luminosity to the radio emission with the same radio efficiency as NS pulsars, the number of WD pulsars which should have been already detected by radio observation can be estimated as
\begin{equation}
10^3 \left( \frac{\alpha \cdot \eta_{WD}}{\eta_{NS}} \right) \left( \frac{\tau_{WD}}{\tau_{NS}} \right) \left( \frac{L_{WD}}{L_{NS}} \right) \sim 10 \left( \frac{\alpha}{0.01} \right).
\end{equation}
Thus $O(10)$ WD pulsars may well be observed as radio pulsar with relatively long period $P \sim 50 \mathrm{sec}$. However, since the efficiency of the radio emission depends on the detailed situation in the polar cap regions, whether WD pulsars have the same efficiency as NS pulsars is highly uncertain at this stage. Other than the electromagnetic emissions, double degenerate WD mergers, which we consider as the origin of WD pulsars, is a promising source of the future gravitational wave observation by LISA~\cite{LISA:2010}. It is very interesting if we get a strong constraint on the event rate of the mergers in our Galaxy by observing the high energy $e^{\pm}$. In this paper, we consider only merged WDs as a source of high energy $e^{\pm}$ emissions. In the single-degenerate binaries, the accretion could induce the rapid rotation of the WDs. These accreting binary systems could also become WD pulsars if they have strong magnetic fields as AE Aquarii. 

\begin{figure}[htbp]
 \begin{center}
  \includegraphics[width=70mm]{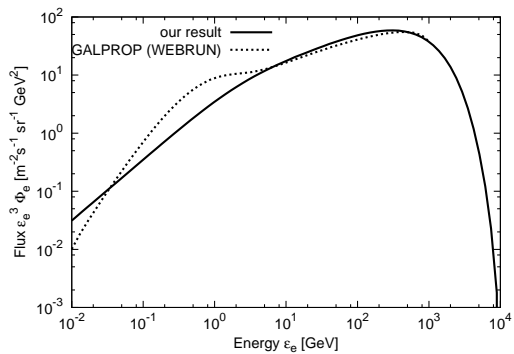}
 \end{center}
 \caption{The comparison of our result and the GALPROP code for $e^{\pm}$ total flux from multiple NS pulsars.}
 \label{test}
\end{figure}

At the current moment, our model have several crucial assumptions which should be considered more carefully. Last of all, we discuss these things and what we should consider in the future works. 
\begin{enumerate}
\item[A.] We have discussed the death line of WD pulsars based on the simplest polar cap model, considering only the curvature radiation for the $e^{\pm}$ pair creation photons and the vacuum polar cap gap in which $\rho =0$. It has been shown that the inverse Compton scattered photon is important for the $e^{\pm}$ pair creation in the polar cap, and the observed death line of NS pulsars is well explained also by the space charge limited flow model~\cite{Harding:2001,Harding:2002}. Especially for the WD, even in the case of ${\bf \Omega} \cdot {\bf B} < 0$, the charged limited flows may exist since the binding energy of the ions could be smaller than the thermal energy at the surface. Hence we have to investigate the death line for WD pulsars based on, for example, the Harding \& Muslimov model~\cite{Harding:2001,Harding:2002}. Also the $e^{\pm}$ multiplicity in the magnetosphere is crucial for the maximum energy, and we have to calculate it consistently with the polar cap model.
\item[B.] There are uncertainties about the accelerating and cooling processes of $e^{\pm}$ in the pulsar wind nebula. The energy flux of the magnetic field may not be conserved in the wind region like the Crab nebula~\cite{Rees:1974,KennelCoroniti:1984}. This time we have assumed Eq.(\ref{Bconf2}) for simplicity. Moreover, we have to evaluate more precisely the inverse Compton scattering in the pulsar wind nebula as a radiative cooling. In this paper we roughly estimate it to be comparable to the synchrotron radiation. We also have to worry about whether the wind mainly consist of the $e^{\pm}$, which is still under debate even in the case of NS pulsars. 
\item[C.] When calculating the $e^{\pm}$ flux from the multiple sources, we assume that the source distribution and the $e^{\pm}$ diffusion process are isotropic. However compact objects like WDs and NSs may distribute more densely near the center of our Galaxy. (Also, the large kick which could be given at their birth may affect the spatial distribution of the pulsars.) Since the arrival anisotropy can be useful to discriminate the origin of the observed $e^{\pm}$, we should take into account these anisotropic effects. For the $e^{\pm}$ with relatively low energy $\lesssim 10 \mathrm{GeV}$, the inverse Compton energy losses become less important and consequently the propagation range of $e^{\pm}$ increases, i.e. the anisotropic effects during the propagation, for example the effect of the Galactic disk structure, become more prominent. In these low energy range, the public GALPROP code~\cite{GALPROP} can provide a more reliable calculation of the propagation from distant sources arbitrarily distributed in our Galaxy. Fig.\ref{test} shows the comparison of our result and the GALPROP code~(WEBRUN~\cite{GALPROP}) for the primary $e^{\pm}$ total flux from multiple NS pulsars with the same parameters as Fig.\ref{model_B}. We have confirmed that our result is consistent with the more realistic calculation in the high energy region and begin to deviate below $\lesssim 10 \mathrm{GeV}$. The bump around $0.5 \mathrm{GeV}$ in the result of the GALPROP code is formed mainly due to the diffusive reacceleration of $e^{\pm}$ during the propagation in our Galaxy. Note that in this region, the solar modulation is relevant and the uncertainty becomes large.
\end{enumerate}

\section{Acknowledgements}
We thank T.~Piran, I.~V.~Moskalenko, Y.~Suwa, Y.~Ohira, K.~Murase, H.~Okawa, F.~Takahara, S.~Shibata and T.~Nakamura for many useful discussions and comments. K.K acknowledge the support of the Grant-in-Aid for the Global COE Program "The Next Generation of Physics, Spun from Universality and Emergence" from the Ministry of Education, Culture, Sports, Science and Technology (MEXT) of Japan. This work is also supported by the Grant-in-Aid from the MEXT of Japan, Nos. 19047004, 21684014, 22244019, 22244030 for K.I. and 22740131 for N.K..

\end{document}